%% file: paper_fde.tex
\documentclass[10pt,journal]{IEEEtran}
\usepackage{ifpdf}
\usepackage{balance} 

\usepackage{color}
\usepackage{url}
\usepackage{booktabs} 
\usepackage{cite}
\usepackage{graphicx}
\usepackage[dvipsnames]{xcolor}
\usepackage[subrefformat=parens, labelformat=parens]{subfig}
\usepackage[font=footnotesize, labelfont=footnotesize]{caption}
\usepackage{caption}
\usepackage{amsmath,amssymb,amsfonts}
\usepackage{multirow,makecell}

\usepackage{siunitx} 
\DeclareSIUnit \MHz{ MHz}
\DeclareSIUnit \pF{ pF}
\DeclareSIUnit \nH{ nH}
\DeclareSIUnit \dB{ dB}
\DeclareSIUnit \dBm{ dBm}
\DeclareSIUnit \deg{ deg}

\usepackage{algorithm, algorithmicx, algpseudocode}

\usepackage{tabularx}
\usepackage{enumitem}

\newcommand{\addedMK}[1]{{\color{black}#1}}

\newif\ifshowedit
\showedittrue

\usepackage[normalem]{ulem}
\ifshowedit
\newcommand{\remove}[1]{\textcolor{red}{{\sout{#1}}}}

\else
\newcommand{\remove}[1]{}

\fi

\begin{document}

\sloppy

\title{Design and Testbed Deployment of Frequency-Domain Equalization Full Duplex Radios}

\author{Manav~Kohli, 
Mahmood~Baraani~Dastjerdi, 
Jin~Zhou, 
Ivan~Seskar, \\
Harish~Krishnaswamy, 
Gil~Zussman,
and~Tingjun~Chen
\vspace{-\baselineskip}
\thanks{This work was supported in part by NSF grants CNS-1827923, EEC-2133516, DGE-2036197, CNS-2148128 and by funds from federal agency and industry partners as specified in the NSF Resilient \& Intelligent NextG Systems (RINGS) program. A partial and preliminary version of this paper appeared in ACM MobiCom'19, Oct. 2019~\cite{chen2019wideband}.}
\thanks{M. Kohli, M. Baraani Dastjerdi, H. Krishnaswamy, and G. Zussman are with the Department of Electrical Engineering, Columbia University, New York, NY, USA (email: \{mpk2138@, b.mahmood@, harish@ee., gil@ee.\}columbia.edu).}
\thanks{J. Zhou is with the Department of Electrical and Computer Engineering, University of Illinois at Urbana-Champaign, Urbana, IL, USA (e-mail: jinzhou@illinois.edu).}
\thanks{I. Seskar is with WINLAB, Rutgers University, New Brunswick, NJ, USA (e-mail: seskar@winlab.rutgers.edu)}
\thanks{T. Chen is with the Department of Electrical and Computer Engineering, Duke University, Durham, NC, USA (e-mail: tingjun.chen@duke.edu).}
}

\setlength{\abovedisplayskip}{6pt}
\setlength{\belowdisplayskip}{6pt}

\maketitle

\begin{abstract}
Full-duplex (FD) wireless can significantly enhance spectrum efficiency but requires effective self-interference (SI) \addedMK{cancellers. RF SI cancellation (SIC) via frequency-domain equalization (FDE), where bandpass filters channelize the SI, is suited for integrated circuits (ICs). In this paper, we explore the limits and higher layer challenges associated with using such cancellers. We evaluate the performance of a custom FDE-based canceller using two testbeds; one with mobile FD radios and the other with upgraded, static FD radios in the PAWR COSMOS testbed. The latter is a lasting artifact for the research community, alongside a dataset containing baseband waveforms captured on the COSMOS FD radios, facilitating FD-related experimentation at the higher networking layers. We evaluate the performance of the FDE-based FD radios in both testbeds, with experiments} showing 95 dB overall achieved SIC (52 dB from RF SIC) across 20 MHz bandwidth, and an average link-level FD rate gain of 1.87$\times$. We also conduct experiments in (i) uplink-downlink networks with inter-user interference, and (ii) heterogeneous networks with half-duplex and FD users. The experimental FD gains in the two types of networks depend on the users’ SNR values and the number of FD users, and are 1.14$\times$--1.25$\times$ and 1.25$\times$--1.73$\times$, respectively, confirming previous analytical results.
\end{abstract}


\begin{IEEEkeywords}
Full-duplex wireless, frequency-domain equalization, self-interference cancellation, software-defined radios\addedMK{, wireless experimentation testbeds}
\end{IEEEkeywords}

\input{tex/nomenclature}

\section{Introduction}
\label{sec:intro}
\input{tex/intro}

\section{Related Work}
\label{sec:related}
\input{tex/related}

\section{Background and Formulation}
\label{sec:background}
\input{tex/background}

\section{Design and Optimization}
\label{sec:impl}
\input{tex/impl}


\section{\addedMK{Experimentation in the Mobile Testbed}}
\label{sec:exp}
\input{tex/experiments_cosmos.tex}



\section{Conclusion}
\label{sec:conclusion}
\input{tex/conclusion}


\bibliographystyle{IEEEtran}
\bibliography{paper_fde}

\end{document}

%% file: tex/nomenclature.tex
\newcommand{\littlesum}{\mathop{\textstyle\sum}}
\newcommand{\littleint}{\mathop{\textstyle\int}}

\newcommand{\SIChnlTF}{H_{\textrm{SI}}}
\newcommand{\SIChnlTFvec}{\mathbf{H}_{\textrm{SI}}}
\newcommand{\AntTF}{H_{\textrm{SI}}}

\newcommand{\NumTap}{M} 
\newcommand{\NumChnl}{K}
\newcommand{\BW}{B}
\newcommand{\RFCancTF}{H}
\newcommand{\RFCancTapTF}[1]{H_{#1}}
\newcommand{\ResTF}{H_{\textrm{res}}}
\newcommand{\ResTapTF}[1]{H_{\textrm{res}, #1}}
\newcommand{\OptProblem}{\textsf{(P1)}}
\newcommand{\OptProblemIC}{\textsf{(P3)}}
\newcommand{\OptProblemPCB}{\textsf{(P2)}}
\newcommand{\OptProblemICPCB}{\textsf{(P4)}}

\newcommand{\FDECancTF}{H^{\textrm{FDE}}}
\newcommand{\FDETapTF}[1]{H_{#1}^{\textrm{FDE}}}

\newcommand{\TF}{H}
\newcommand{\TFApprox}{\widetilde{H}}
\newcommand{\ICTF}{H^{\textrm{I}}}
\newcommand{\ICResTF}{H_{\textrm{res}}^{\textrm{I}}}
\newcommand{\ICTapTF}[1]{H_{#1}^{\textrm{I}}}
\newcommand{\ICTapTFApprox}[1]{\widetilde{H}_{#1}^{\textrm{I}}}
\newcommand{\ICTapAmp}[1]{A_{#1}^{\textrm{I}}}
\newcommand{\ICTapAmpMin}{A_{\textrm{min}}^{\textrm{I}}}
\newcommand{\ICTapAmpMax}{A_{\textrm{max}}^{\textrm{I}}}
\newcommand{\ICTapPhase}[1]{\phi_{#1}^{\textrm{I}}}
\newcommand{\ICTapCF}[1]{f_{\textrm{c},#1}}
\newcommand{\ICTapCFMin}{f_{\textrm{c,min}}}
\newcommand{\ICTapCFMax}{f_{\textrm{c,max}}}
\newcommand{\ICTapCFNoidx}{f_{\textrm{c}}}
\newcommand{\ICTapQF}[1]{Q_{#1}}
\newcommand{\ICTapQFMin}{Q_{\textrm{min}}}
\newcommand{\ICTapQFMax}{Q_{\textrm{max}}}
\newcommand{\PCBTF}{H^{\textrm{P}}}
\newcommand{\PCBResTF}{H_{\textrm{res}}^{\textrm{P}}}
\newcommand{\PCBTapTF}[1]{H_{#1}^{\textrm{P}}}
\newcommand{\BPFTapTF}[1]{H_{#1}^{\textrm{B}}}
\newcommand{\PCBTapTFApprox}[1]{\widetilde{H}_{#1}^{\textrm{P}}}
\newcommand{\PCBTapAmp}[1]{A_{#1}^{\textrm{P}}}
\newcommand{\PCBTapAmpMin}{A_{\textrm{min}}^{\textrm{P}}}
\newcommand{\PCBTapAmpMax}{A_{\textrm{max}}^{\textrm{P}}}
\newcommand{\PCBTapPhase}[1]{\phi_{#1}^{\textrm{P}}}
\newcommand{\PCBTapCF}[1]{f_{c,#1}^{\textrm{P}}}
\newcommand{\PCBTapCFCapNoidx}{C_{\textrm{F}}}
\newcommand{\PCBTapCFCap}[1]{C_{\textrm{F},#1}}
\newcommand{\PCBTapCFCapMin}{C_{\textrm{F,min}}}
\newcommand{\PCBTapCFCapMax}{C_{\textrm{F,max}}}
\newcommand{\PCBTapQFCap}[1]{C_{\textrm{Q},#1}}
\newcommand{\PCBTapQFCapNoidx}{C_{\textrm{Q}}}
\newcommand{\PCBTapQFCapMin}{C_{\textrm{Q,min}}}
\newcommand{\PCBTapQFCapMax}{C_{\textrm{Q,max}}}
\newcommand{\ICPCBTF}{H^{\textrm{I/P}}}
\newcommand{\ICPCBTapTF}[1]{H_{#1}^{\textrm{I/P}}}

\newcommand{\CenterFreq}[1]{f_{c,#1}}
\newcommand{\CenterFreqSingle}{f_c}
\newcommand{\CenterFreqVec}{\mathbf{f}_{c}}
\newcommand{\QFactor}[1]{Q_{#1}}
\newcommand{\QFactorVec}{\mathbf{Q}}
\newcommand{\Amp}[1]{A_{#1}}
\newcommand{\AmpVec}{\mathbf{A}}
\newcommand{\Phase}[1]{\phi_{#1}}
\newcommand{\PhaseVec}{\bm{\upphi}}
\newcommand{\IterSICFreq}[1]{f_{#1}^{\textrm{SIC}}}
\newcommand{\IterSICFreqVec}{\mathbf{f}^{\textrm{SIC}}}
\newcommand{\CF}[1]{f_{c,#1}}
\newcommand{\CFNoidx}{f_c}
\newcommand{\QF}[1]{Q}

\newcommand{\CapCF}[1]{C_F}
\newcommand{\CapCFVec}{\mathbf{C}_F}
\newcommand{\CapQF}[1]{C_Q}
\newcommand{\CapQFVec}{\mathbf{C}_Q}
\newcommand{\Att}[1]{A_{#1}}
\newcommand{\AttVec}{\mathbf{A}}
\newcommand{\PS}[1]{\phi_{#1}}
\newcommand{\PSVec}{\bm{\upphi}}

\newcommand{\PtxNoIdx}{P_{\textrm{tx}}}
\newcommand{\Ptx}[1]{P_{\textrm{tx},{#1}}}
\newcommand{\Prx}[1]{P_{\textrm{rx},{#1}}}
\newcommand{\Dist}[1]{d_{#1}}
\newcommand{\PathGain}{G}
\newcommand{\Nrx}{N_{\textrm{rx}}}
\newcommand{\SNR}[1]{\gamma_{#1}}
\newcommand{\SNRUL}{\gamma_{\textrm{UL}}}
\newcommand{\SNRDL}{\gamma_{\textrm{DL}}}
\newcommand{\XINR}{\gamma_{\textrm{Self}}}
\newcommand{\IUI}{\gamma_{\textrm{IUI}}}
\newcommand{\DataRate}[1]{r_{#1}}

\newcommand{\MyNormTwo}[1]{\left| #1 \right|_2}
\newcommand{\NormTwo}[1]{\left| #1 \right|}


%% file: tex/intro.tex
Full-duplex (FD) wireless -- simultaneous transmission and reception on the same frequency channel -- can significantly improve spectrum efficiency at the physical (PHY) layer and provide many other benefits at the higher layers~\cite{sabharwal2014band,krishnaswamy2016spectrum,zhang2016fullduplex}. The main challenge associated with FD is the extremely strong self-interference (SI) signal that needs to be suppressed, requiring {80--110}\thinspace{dB} of SI cancellation (SIC). 

\addedMK{Prior work} leveraging off-the-shelf components and software-defined radios (SDRs) has established the feasibility of FD wireless through SI suppression at the antenna interface, and SIC in analog/RF and digital domains~\cite{choi2010achieving,duarte2010full,bharadia2013full,korpi2016full,chen2021survey}. However, RF cancellers achieving wideband SIC \addedMK{at sub-6\thinspace{GHz} operating frequencies} (e.g.,~\cite{bharadia2013full, korpi2016full}) \addedMK{commonly rely} on transmission-line delays, which cannot be realized in small-form-factor nodes and/or integrated circuits (ICs) due to the required length for generating nanosecond-scale time delays and the lossy nature of the silicon substrate.\footnote{For instance, obtaining a nanosecond delay in silicon typically requires a {15}\thinspace{cm}-long delay line.}

\begin{figure}[!t]
\centering
\vspace{-0.5\baselineskip}
\subfloat[]{
\label{fig:intro-pcb}
\includegraphics[height=1.05in]{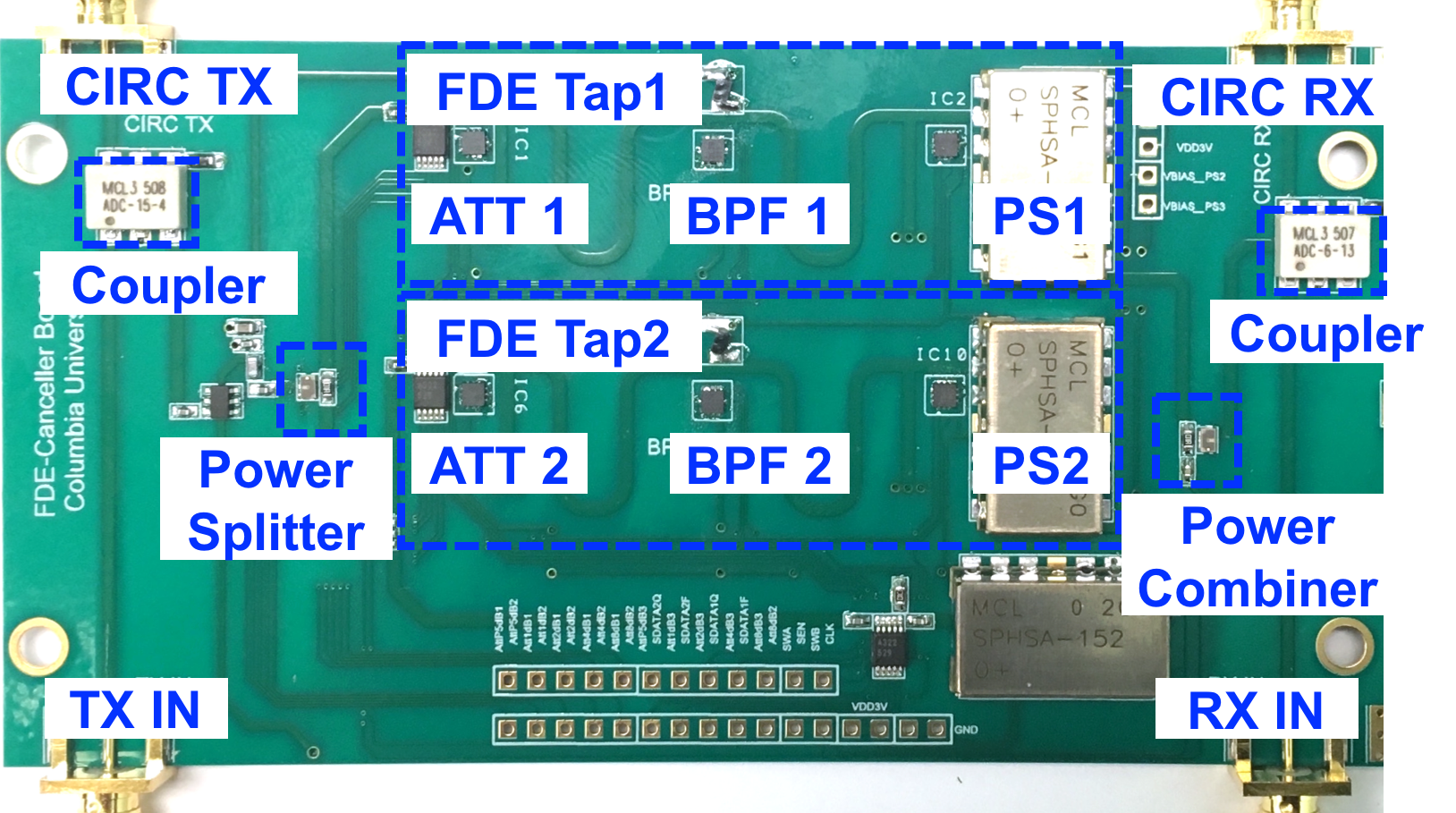}}
\hspace{-6pt} \hfill
\subfloat[]{
\label{fig:intro-fd-radio}
\includegraphics[height=1.05in]{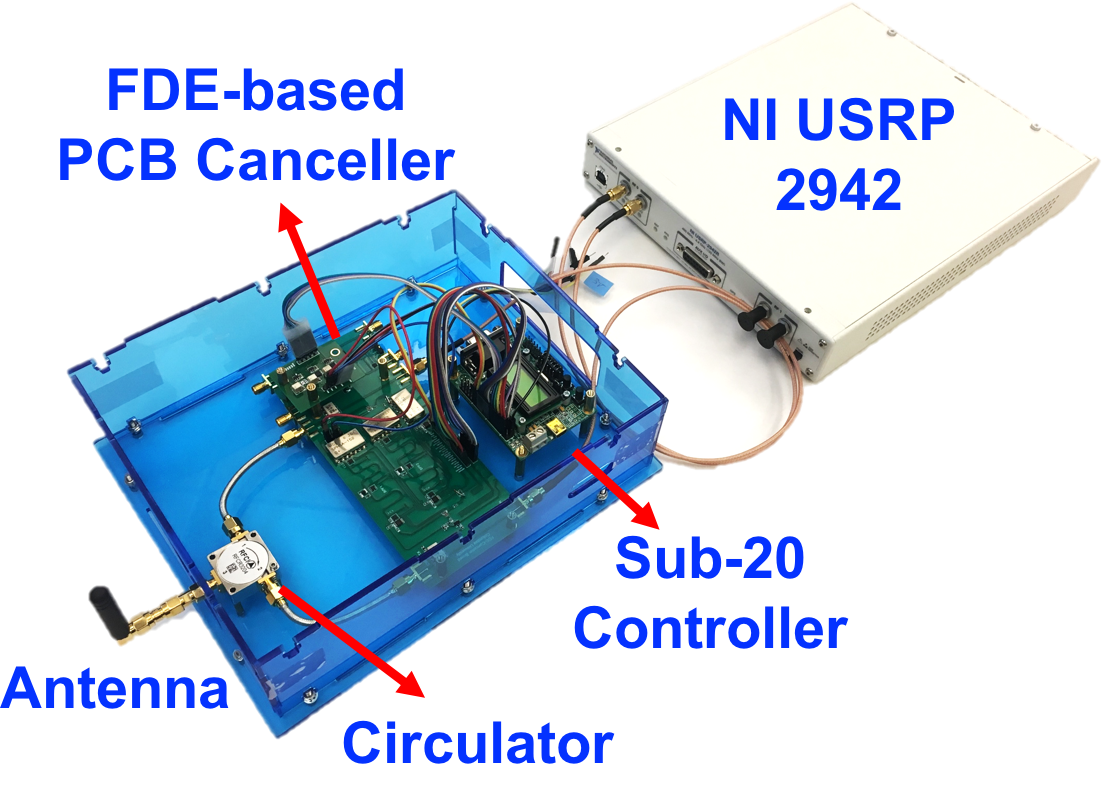}}
\\ \vspace{-.75\baselineskip}
\subfloat[]{
\label{fig:intro-fd-net}
\includegraphics[width=0.95\columnwidth]{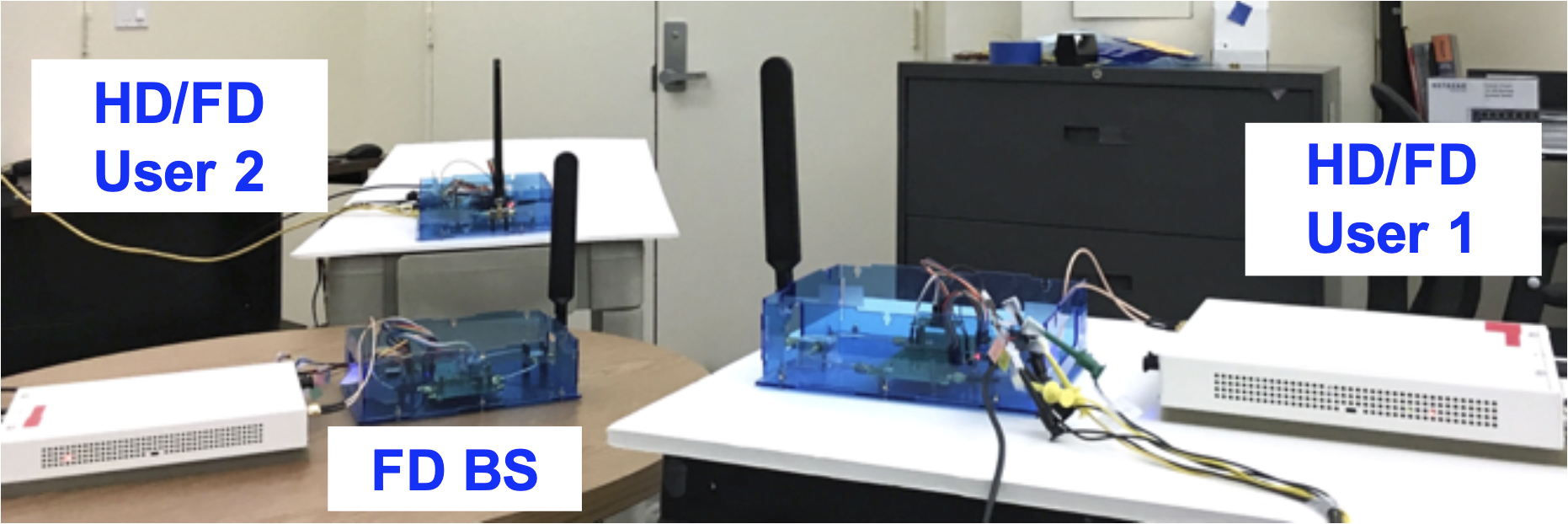}}
\vspace{-0.5\baselineskip}
\caption{(a) The frequency-domain equalization- (FDE-) based wideband RF canceller implemented using discrete components on a printed circuit board (PCB), (b) the implemented FDE-based \addedMK{mobile} full-duplex (FD) radio, and (c) the \addedMK{mobile testbed} consisting of an FD base station (BS) and 2 users \addedMK{mounted on carts} that can operate in either half-duplex (HD) or FD mode.}
\label{fig:intro}
\vspace{-\baselineskip}
\end{figure}

\begin{figure}[!t]
\centering
\subfloat[]{
\includegraphics[width=0.315\columnwidth]{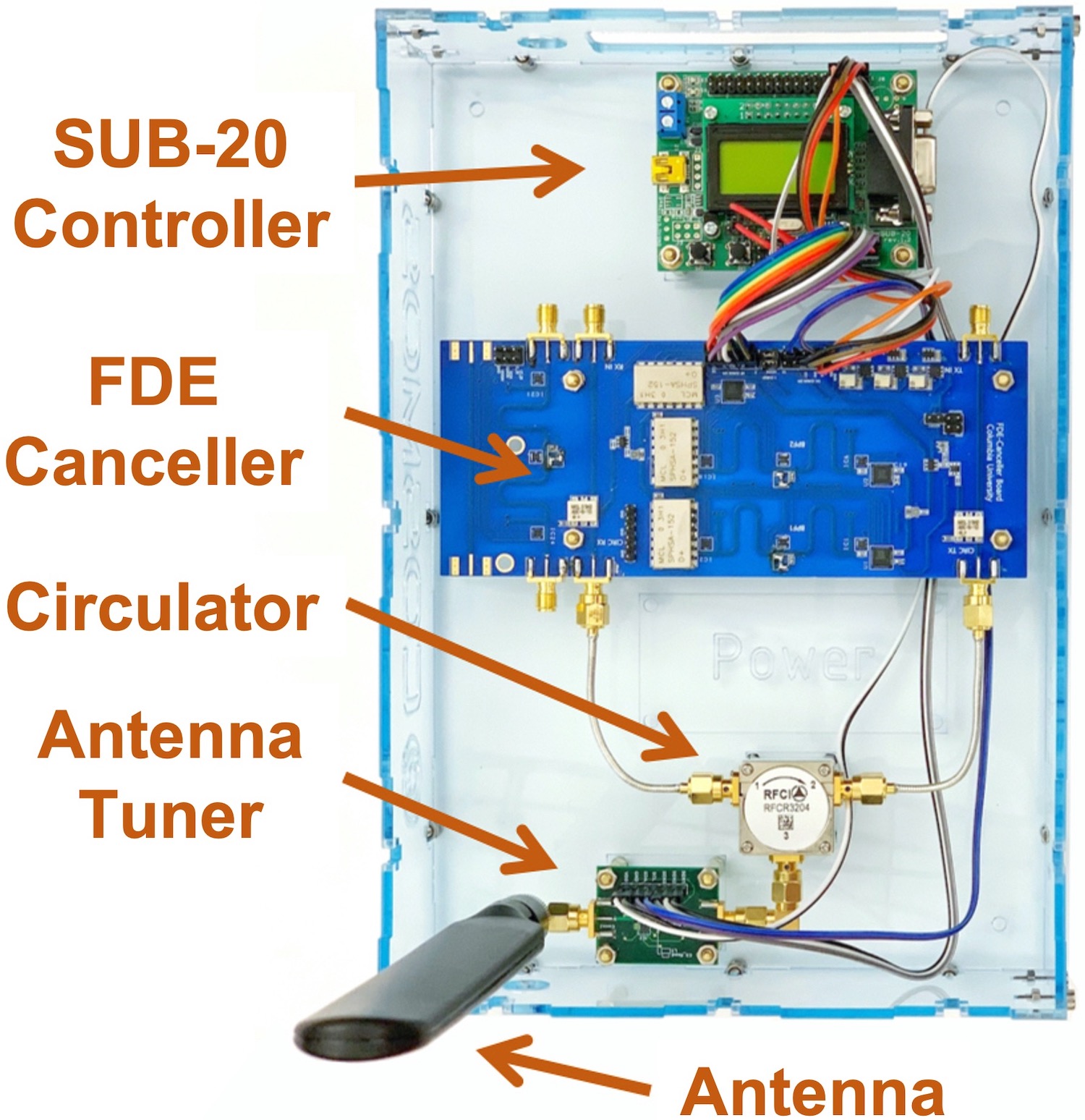}
\label{fig:gen2-integration-box}
}
\subfloat[]{
\includegraphics[width=0.632\columnwidth]{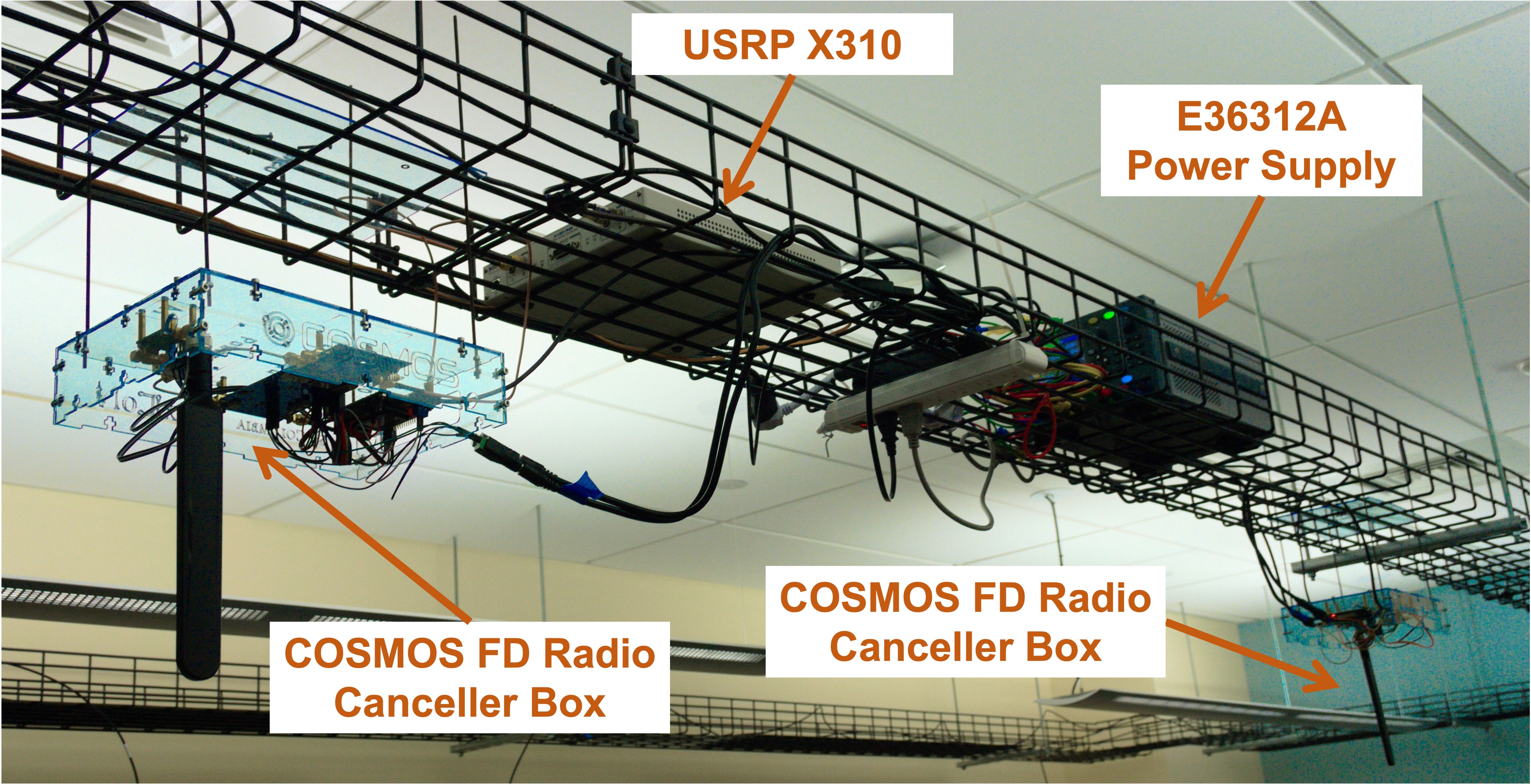}
\label{fig:gen2-integration-rack}
}
\vspace{-0.5\baselineskip}
\caption{\addedMK{Integration of the FDE-based FD radios in COSMOS Sandbox 2. (a) labelled diagram of the ``canceller box", showing the various system components. (b) Two out of the four canceller boxes mounted in the testbed.}}
\label{fig:gen2-integration}
\end{figure}

A \emph{compact IC-based} design is necessary for supporting FD in hand-held devices (e.g., handsets and tablets)~\cite{yang2015wideband,Zhou_WBSIC_JSSC15,krishnaswamy2016spectrum,zhou2017integrated,yi2020inband,wang2022fullduplex,chen2021survey}. \addedMK{There are several approaches at the circuit-level enabling wideband RF SIC on RFICs~\cite{zhang2016fullduplex,chen2021survey} including switched-capacitor delay lines~\cite{wang2022fullduplex,nagulu2021fullduplex}, Tx-Rx isolation via bidirectional frequency conversion~\cite{yi2020inband}, and frequency-domain equalization (FDE)~\cite{garimella2023fde, Zhou_WBSIC_JSSC15}. In this paper, we focus on the FDE method initially presented in~\cite{Zhou_WBSIC_JSSC15}, which in contrast to the delay line-based approaches that essentially perform time-domain equalization, utilizes} tunable, reconfigurable, high quality factor 2$^{\textrm{nd}}$-order bandpass filters (BPFs) with amplitude and phase controls to emulate the frequency-selective antenna interface. In general, tunable, high quality factor BPFs are perhaps as hard to implement on an IC as nanosecond-scale delay lines. \addedMK{However, their implementation in nanoscale CMOS has been enabled by advances in $N$-path filters~\cite{reiskarimian2016analysis}.}

While major advances have been made at the \addedMK{RFIC} level, \addedMK{existing work has several limitations: (i) the limits of achievable RF SIC for FDE-based FD systems} have not been fully understood, (ii) \addedMK{RF canceller configuration schemes} need to be developed in order to achieve optimized and adaptive RF SIC in real-world environments, and (iii) the system-level performance of such IC-based FD radios has not been evaluated in different network settings. Therefore, in this paper we focus on \addedMK{the evaluation of FDE-based RF cancellers in two experimental testbeds: a lab-based \textit{mobile FD testbed} (Fig.~\ref{fig:intro}) and the \textit{COSMOS FD testbed} (Fig.~\ref{fig:gen2-integration}).}

\addedMK{Since interfacing an RFIC canceller to an SDR presents numerous technical challenges, we design and implement two iterations of a printed circuit board (PCB) emulation of the FDE-based RFIC canceller presented in~\cite{Zhou_WBSIC_JSSC15} with discrete components on a printed circuit board (PCB). The first iteration of the PCB canceller appears in the \textit{mobile FD radios} shown in Fig.~\ref{fig:intro}. The second, more robust iteration of the PCB canceller has been integrated in the PAWR COSMOS testbed~\cite{raychaudhuri2020challenge}, shown in Fig.~\ref{fig:gen2-integration}. The \textit{COSMOS FD radios} are intended to be a lasting artifact for the research community, as they are openly and remotely accessible for experimentation~\cite{fd_tutorial_wiki}. Both testbeds facilitate experimentation using SDRs in a network with multiple FD radios, while the mobile testbed allows for evaluation of the canceller configuration scheme.}

\addedMK{The COSMOS FD testbed improves on the first such effort by the Columbia FlexICoN project~\cite{flexicon} using the ORBIT testbed~\cite{flexicon_orbit_gen1,kohli2021open}. Specifically, the COSMOS FD testbed contains four FDE-based FD radios operating at 4$\times$ the bandwidth as ORBIT's single narrowband FD radio~\cite{Zhou_NCSIC_JSSC14}. The COSMOS FD testbed has been improved from its prior iteration~\cite{kohli2021open} through the uniform use of higher performance SDRs for each FD radio and integration with the COSMOS servers, which provide significantly greater compute capability. Additionally, we provide a publicly available dataset containing baseband I/Q samples for OFDM-modulated packets.}

\addedMK{We create the mobile FD radios by integrating the PCB canceller with an NI USRP SDR, as depicted in Fig.~\ref{fig:intro}\subref{fig:intro-fd-radio}. A similar procedure is used for the COSMOS FD radios, shown in Fig.~\ref{fig:gen2-integration}\subref{fig:gen2-integration-rack}.}
\addedMK{Both the mobile and COSMOS FD radios achieve {85--95}\thinspace{dB} overall SIC across {20}\thinspace{MHz} real-time bandwidth, enabling an FD link budget of {0--10}\thinspace{dBm} average TX power level and $-${85}\thinspace{dBm} RX noise floor.} In particular, \addedMK{up to} {52}\thinspace{dB} RF SIC is achieved, from which {20}\thinspace{dB} is obtained from the antenna interface isolation. We present a realistic model of the PCB canceller, \addedMK{and use it to develop a configuration scheme based on an optimization problem for the mobile FD radios, allowing efficient adaption of the canceller to environmental changes.} The PCB canceller model is experimentally validated and is shown to have high accuracy. 


\addedMK{Using both testbeds}, we extensively evaluate the network-level FD gain and confirm analytical results in two types of networks: (i) \emph{UL-DL networks} consisting of one FD base station (BS) and two half-duplex (HD) users with inter-user interference (IUI), and (ii) \emph{heterogeneous HD-FD networks} consisting of one FD BS and co-existing HD and FD users. \addedMK{For the UL-DL network on the mobile testbed}, we show experimentally that the throughput gain is between  1.14$\times$--1.25$\times$ compared to 1.22$\times$--1.30$\times$ predicted by analysis. 
For heterogeneous HD-FD networks, we demonstrate the impact of different user SNR values and the number of FD users on both the FD gain and the fairness performance measured by the Jain's fairness index.
For example, in a 4-node network consisting of an FD BS and 3 users with various user locations and SNR values, the median network throughout can be improved by 1.25$\times$ and 1.52$\times$ when one and two users become FD-capable, respectively. \addedMK{The COSMOS FD radios provide very similar results for these two types of networks under appropriately defined experimental conditions.}

To the best of our knowledge, this is the first experimental study of \addedMK{FDE-based FD radios} in such networks using \addedMK{testbeds comprised} of both HD and FD radios. The results demonstrate the practicality and \addedMK{validate the} performance of FDE-based FD radios, which are suitable for small-form-factor devices. The \addedMK{presented results, as well as openly-accessible COSMOS FD radios,} can also serve as building blocks for developing higher layer (e.g., MAC) protocols. 


To summarize, the main contributions of the paper are:
\begin{enumerate}[leftmargin=*,topsep=0pt] 
\item[1.]
We present the design, implementation, modeling, and validation of the FDE-based PCB canceller, as well as an optimized canceller configuration scheme;
\item[2.]
\addedMK{We experimentally evaluate the FD throughput gain in various network settings with different user capabilities (i.e., HD or FD) and user SNR values.}

\item[3.]
\addedMK{We provide the open-access COSMOS FD testbed radios and dataset, and provide results to show the effectiveness of the testbed for network-level experimentation.} 
\end{enumerate}

The rest of the paper is organized as follows. Section~\ref{sec:related} reviews related work. In Section~\ref{sec:background}, we present the problem formulation for the FDE method. We present the design, implementation, and model of the FDE-based PCB canceller, as well as the optimized canceller configuration scheme in Section~\ref{sec:impl}. 
\addedMK{The performance of FD radios based on the FDE PCB cancellers are experimentally evaluated in the lab and COSMOS testbeds in Sections~\ref{sec:exp} and \ref{ssec:cosmos}. We also provide information about the publicly available RF SIC waveform dataset in Section~\ref{ssec:cosmos}.} 
\addedMK{Lastly,} we conclude and discuss future directions in Section~\ref{sec:conclusion}.


%% file: tex/related.tex
Extensive research related to FD wireless is summarized \addedMK{in~\cite{chen2021survey, kolodziej2019survey, sabharwal2014band},} including implementations of FD radios and systems, analysis of rate gains, and resource allocation at the higher layers. Below, we briefly review the related work.

\noindent\textbf{RF canceller and FD radio designs}.
RF SIC typically involves two stages: (i) isolation at the antenna interface, and (ii) SIC in the RF domain using cancellation circuitry. While a separate TX/RX antenna pair can provide good isolation and can be used to achieve cancellation~\cite{radunovic2010rethinking,choi2010achieving,jain2011practical,aryafar2012midu}, a shared antenna interface such as a circulator is more appropriate for single-antenna implementations~\cite{bharadia2013full,chung2015prototyping} and is compatible with FD MIMO systems. Existing designs of analog/RF SIC circuitry are mostly based on a time-domain interpolation approach~\cite{bharadia2013full,korpi2016full}. 
\addedMK{Several FD MIMO radio designs based on this approach} are presented~\cite{aryafar2012midu,duarte2013design,bharadia2014full,chen2015flexradio,chung2017compact}. FD relays have also been successfully demonstrated in~\cite{bharadia2015fastforward,chen2017bipass}. Moreover, SIC can be achieved via digital/analog beamforming in FD massive-antenna systems~\cite{everett2016softnull,aryafar2018pafd}.
The techniques utilized in these works are incompatible with IC \addedMK{implementations required} for small-form-factor devices.
In this paper, we focus on an FDE-based canceller, which builds on our previous work towards the design of such an RFIC canceller~\cite{Zhou_WBSIC_JSSC15}. However, existing IC-based FD radios \addedMK{(e.g.,~\cite{Zhou_WBSIC_JSSC15,wang2022fullduplex,yi2020inband})} have not been evaluated at the system-level in different network settings. \addedMK{FD has been studied in the context of millimeter-wave radios, with SIC performed across antenna, RF, and digital domains~\cite{singhmmwavefd2020,MITRadar140G,yu2023realizing}. Lastly, the benefits of FD to beyond-5G and 6G networks has been explored~\cite{smida2023fullduplex,harada2022development}.}

\noindent\textbf{FD gain at the link- and network-level}.
At the higher layers, recent work focuses on characterizing the capacity region and rate gains, as well as developing resource allocation algorithms under both perfect~\cite{ahmed2013rate,Sabharwal_DistributedSideChannel_TWC13} and imperfect SIC~\cite{goyal2015full,marasevic2017resource,diakonikolas2017rate}\addedMK{, as well as various levels of available channel state information~\cite{braga2022rate}.} Similar problems are considered in FD multi-antenna/MIMO systems\addedMK{~\cite{zheng2015joint,everett2016softnull,qian2017concurrent, singh2020fullduplex, islam2022rate}}. Medium access control (MAC) algorithms are studied in networks with all HD users~\cite{chen2017probabilistic}, with all FD users~\cite{doost2015performance}, or with heterogeneous HD and FD users~\cite{chen2018hybrid}. Moreover, network-level FD gain is analyzed in~\cite{radunovic2010rethinking,xie2014does,qin2016impact,wang2017fundamental} and experimentally evaluated in~\cite{jain2011practical,kim2013janus} where all the users are HD or FD. \addedMK{~\cite{alkhrijah2023multi} proposes a FD MAC algorithm that uses an out-of-band control plane, with data transmitted at 60\thinspace{GHz} mmWave.} Finally,~\cite{hsu2017inter} proposes a scheme to suppress IUI using an emulated FD radio. FD also facilitates different applications including improved PHY layer security~\cite{vo2013counter} and localization~\cite{liu2017network}.

To the best of our knowledge, this is \emph{the first thorough study of wideband RF SIC achieved via a frequency-domain-based approach (which is suitable for compact implementations) that is grounded in real-world implementation and includes extensive system- and network-level experimentation}.


%% file: tex/background.tex
\begin{figure}[!t]
\centering
\includegraphics[width=0.9\columnwidth]{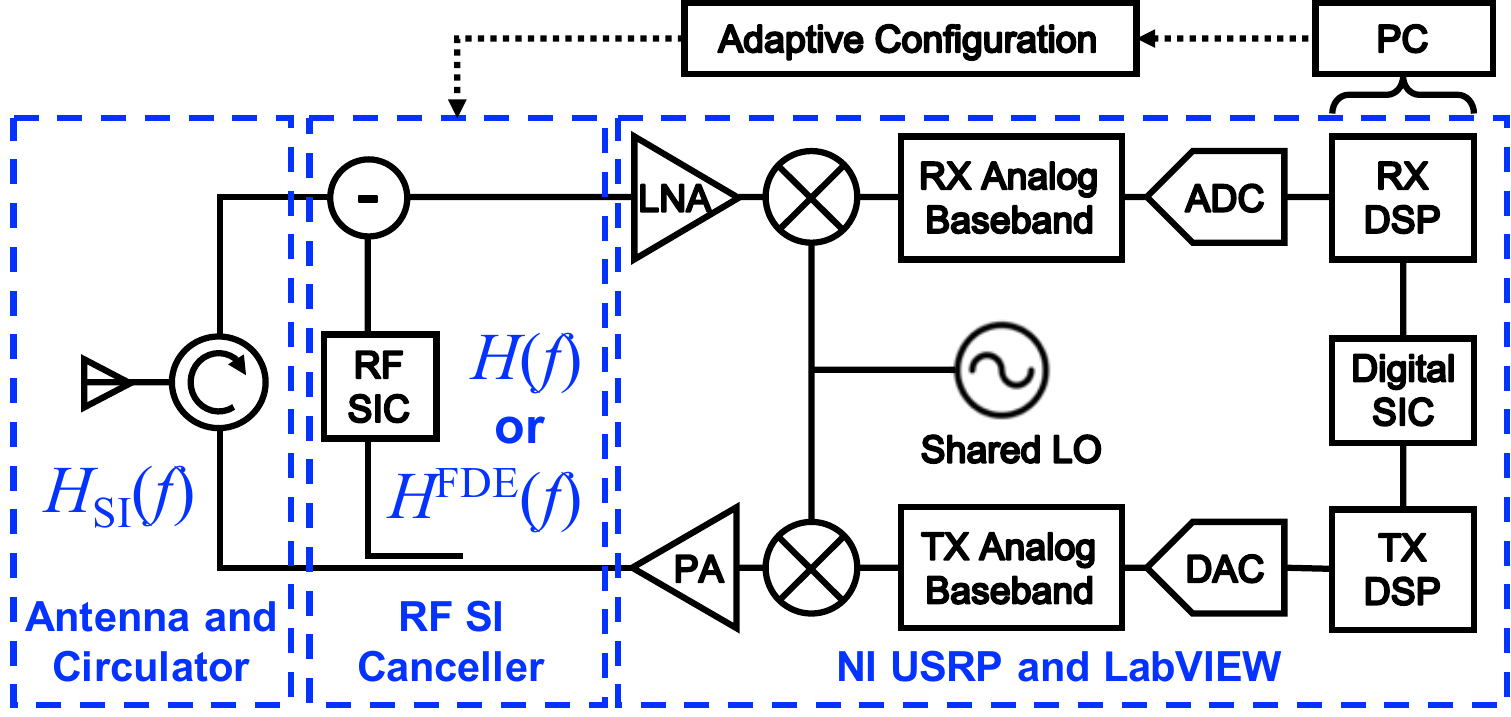}
\caption{Block diagram of an FD radio.}
\label{fig:diagram}
\vspace{-\baselineskip}
\end{figure}

In this section, we briefly review concepts related to FD wireless and RF canceller configuration optimization in the context of an FDE-based canceller. A summary of the main notation is provided in Table~\ref{table:notation}.

\subsection{FD Background and Notation}
Fig.~\ref{fig:diagram} shows the block diagram of a single-antenna FD radio using a circulator at the antenna interface. Due to the extremely strong SI power level and the limited dynamic range of the analog-to-digital converter (ADC) at the RX, a total amount of {80--110}\thinspace{dB} SIC must be achieved across the antenna, RF, and digital domains. Specifically, (i) SI suppression is first performed at the antenna interface, (ii) an RF SI canceller then taps a reference signal at the output of the TX power amplifier (PA) and performs SIC at the input of the low-noise amplifier (LNA) at the RX, and (iii) a digital SI canceller further suppresses the residual SI.

\input{tex/notation}

Consider a wireless bandwidth of $\BW$ that is divided into $\NumChnl$ orthogonal frequency channels. The channels are indexed by $k \in \{1,\dots,\NumChnl\}$ and denote the center frequency of the $k^{\textrm{th}}$ channel by $f_k$.\footnote{We use discrete frequency values $\{f_k\}$ since in practical systems, the antenna interface response is measured at discrete points (e.g., per OFDM subcarrier). However, the presented model can also be applied to cases with continuous frequency values.}
We denote the antenna interface response by $\AntTF(f_k)$ with amplitude $|\AntTF(f_k)|$ and phase $\angle\AntTF(f_k)$. Note that the actual SI channel includes the TX-RX leakage from the antenna interface as well as the TX and RX transfer functions at the baseband from the perspective of the digital canceller. Since the paper focuses on achieving wideband RF SIC, we use $\AntTF(f_k)$ to denote the antenna interface response and also refer to it as the \emph{SI channel}. We refer to \emph{TX/RX isolation} as the ratio (in {dB}, usually a negative value) between the residual SI power at the RX input and the TX output power, which includes the amount of TX/RX isolation achieved by both the antenna interface and the RF canceller/circuitry. We then refer to \emph{RF SIC} as the absolute value of the TX/RX isolation. We also refer to \emph{overall SIC} as the total amount of SIC achieved in both the RF and digital domains. The antenna interface used in our experiments typically provides a TX/RX isolation of around $-${20}\thinspace{dB}.

\subsection{Problem Formulation}
Ideally, an RF canceller is designed to best emulate the antenna interface, $\AntTF(f_k)$, across a desired bandwidth, $\BW=[f_1, f_\NumChnl]$. We denote by $\RFCancTF(f_k)$ the frequency response of an RF canceller and by $\ResTF(f_k) := \AntTF(f_k) - \RFCancTF(f_k)$ the \emph{residual SI channel response}. The optimized RF canceller configuration is obtained by solving {\OptProblem}:
\begin{align}
\textsf{(P1)}\
& \textrm{min:}\
\littlesum\limits_{k=1}^{\NumChnl} \NormTwo{ \ResTF(f_k) }^2 = \littlesum\limits_{k=1}^{\NumChnl} \NormTwo{ \AntTF(f_k) - \RFCancTF(f_k) }^2 \nonumber \\
\textrm{s.t.:}\ & \textrm{constraints on configuration parameters of}\ \RFCancTF(f_k),\ \forall k. \nonumber
\end{align}

The RF canceller configuration obtained by solving $\OptProblem$ minimizes the residual SI power referred to the TX output. As described in Section~\ref{sec:intro}, one main challenge associated with the design of the RF canceller with response $\RFCancTF(f_k)$ to achieve wideband SIC is due to the highly frequency-selective antenna interface, $\AntTF(f_k)$. Moreover, an efficient RF canceller configuration scheme needs to be designed so that the canceller can adapt to time-varying $\AntTF(f_k)$.

\subsection{\addedMK{Overview of an FDE-Based RF Canceller}}
\label{ssec:background-previous-work}

\begin{figure}[!t]
\centering
\vspace{-\baselineskip}
\subfloat[]{
\label{fig:fde-concept-diagram}
\includegraphics[height=1.15in]{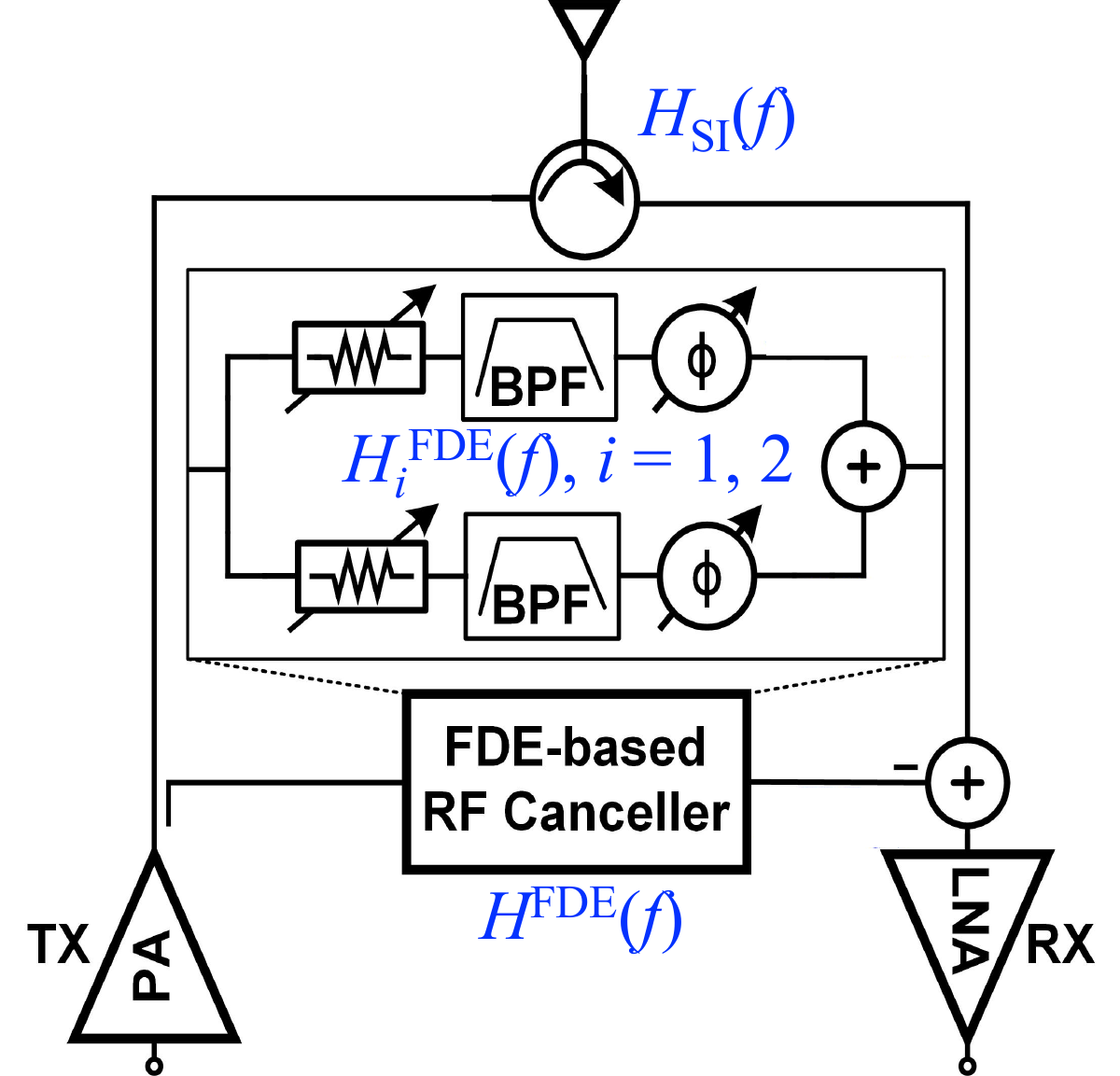}}
\hspace{-12pt} \hfill
\subfloat[]{
\label{fig:fde-concept-bpf}
\includegraphics[height=1.15in]{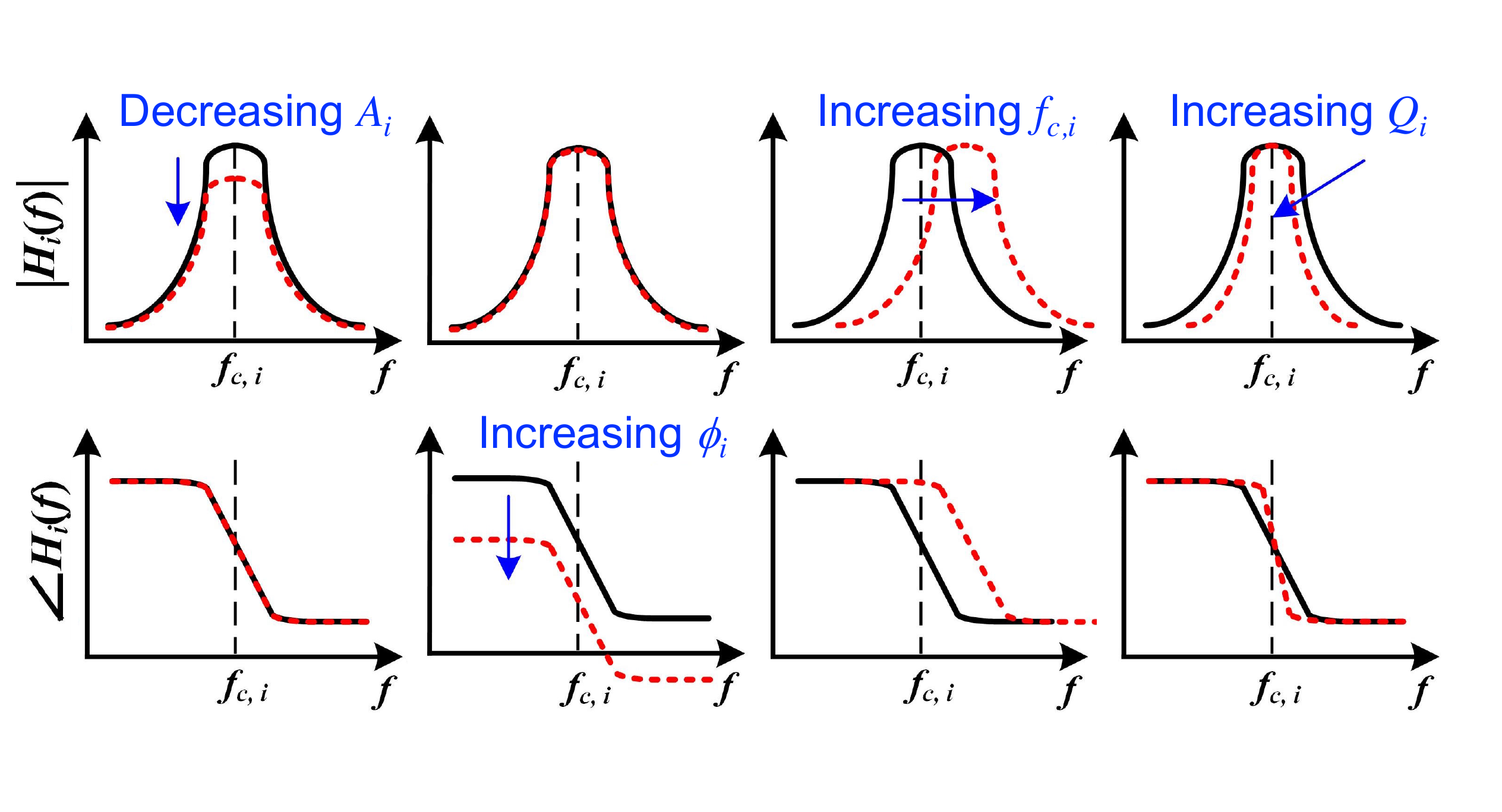}}
\vspace{-0.5\baselineskip}
\caption{(a) Block diagram of an FDE-based RF canceller with $\NumTap=2$ FDE taps, and (b) illustration of amplitude and phase responses of an ideal $2^{\textrm{nd}}$-order bandpass filter (BPF) with amplitude, phase, center frequency, and quality factor (i.e., group delay) controls.}
\label{fig:fde-concept}
\vspace{-\baselineskip}
\end{figure}

Fig.~\ref{fig:fde-concept}\subref{fig:fde-concept-diagram} shows the diagram an \addedMK{FDE-based canceller based on~\cite{Zhou_WBSIC_JSSC15}}, where parallel reconfigurable bandpass filters (BPFs) are used to emulate the antenna interface response across wide bandwidth. We denote the frequency response of a general FDE-based RF canceller consisting of $\NumTap$ FDE taps by
\begin{align}
\label{eq:fde-canc-tf}
\FDECancTF(f_k) & = \littlesum\limits_{i=1}^{\NumTap} \FDETapTF{i}(f_k),
\end{align}
where $\FDETapTF{i}(f_k)$ is the frequency response of the $i^{\textrm{th}}$ FDE tap containing a reconfigurable BPF with amplitude and phase controls. Theoretically, any $m^{\textrm{th}}$-order RF BPF ($m=1,2,\dots$) can be used. Fig.~\ref{fig:fde-concept}\subref{fig:fde-concept-bpf} illustrates the amplitude and phase of a 2$^{\textrm{nd}}$-order BPF with different control parameters. For example, increased BPF quality factors result in ``sharper'' BPF amplitudes and increased group delay.
Since it is shown~\cite{Zhou_WBSIC_JSSC15} that a 2$^{\textrm{nd}}$-order BPF can accurately model the FDE $N$-path filter, the frequency response of an FDE-based RF\underline{I}C canceller with $\NumTap$ FDE taps is given by
\begin{align}
\label{eq:rfic-tf}
\hspace*{-8pt}
\ICTF(f_k) = \littlesum_{i=1}^{\NumTap} \ICTapTF{i}(f_k)
= \littlesum_{i=1}^{\NumTap} \frac{\ICTapAmp{i} \cdot e^{-j\ICTapPhase{i}}}{1 - j\ICTapQF{i} \cdot \left( \ICTapCF{i}/f_k-f_k/\ICTapCF{i}\right)}.
\end{align} 
Within the $i^{\textrm{th}}$ FDE tap, $\ICTapTF{i}(f_k)$, $\ICTapAmp{i}$ and $\ICTapPhase{i}$ are the amplitude and phase controls, and $\ICTapCF{i}$ and $\ICTapQF{i}$ are the center frequency and quality factor of the $2^{\textrm{nd}}$-order BPF (see Fig.~\ref{fig:fde-concept}\subref{fig:fde-concept-bpf}). In the RFIC canceller, $\ICTapCF{i}$ and $\ICTapQF{i}$ are adjusted through a reconfigurable baseband capacitor and transconductors, respectively.

As Fig.~\ref{fig:fde-concept}\subref{fig:fde-concept-bpf} and {\eqref{eq:rfic-tf}} suggest, one FDE tap features four degrees of freedom (DoF) so that the antenna interface, $\AntTF(f_k)$, can be emulated \emph{not only in amplitude and phase, but also in the slope of amplitude and the slope of phase (i.e., group delay)}. This significantly enhances the achievable RF SIC bandwidth.


%% file: tex/notation.tex
\begin{table}[!t]
\caption{Nomenclature}
\label{table:notation}
\vspace{-0.5\baselineskip}
\scriptsize
\begin{center}
\begin{tabular}{| p{0.15\columnwidth} | p{0.75\columnwidth} |}
\hline
$|z|$, $\angle z$ & Amplitude and phase of a complex value $z=x+jy$ ($x,y\in\mathbb{R}$), where $|z| = \sqrt{x^2+y^2}$ and $\angle z = \tan^{-1} \left(\frac{y}{x}\right)$ \\
$\BW$ & Total wireless bandwidth/desired RF SIC bandwidth \\
$\NumChnl$, $k$ & Total number of frequency channels and channel index \\
$f_k$ & Center frequency of the $k^{\textrm{th}}$ frequency channel \\
$\NumTap$ & Number of FDE taps in an FDE-based RF canceller \\
$\AntTF(f_k)$ & Frequency response of the antenna interface \\
$\PCBTF(f_k)$ & Frequency response of the FDE-based PCB canceller \\
$\PCBTapTF{i}(f_k)$  & Frequency response of the $i^{\textrm{th}}$ FDE tap in the PCB canceller \\
$\PCBTapAmp{i}$, $\PCBTapPhase{i}$ & Amplitude and phase controls of the $i^{\textrm{th}}$ FDE tap in the PCB canceller \\
$\PCBTapCFCap{i}$, $\PCBTapQFCap{i}$ & Digitally tunable capacitors that control the center frequency and quality factor of the $i^{\textrm{th}}$ FDE tap in the PCB canceller \\
\hline
\end{tabular}
\end{center}
\vspace{-0.5\baselineskip}
\end{table}


%% file: tex/impl.tex
In this section, we present our design and implementation of an FDE-based canceller using discrete components on a PCB (referred to as the \emph{PCB canceller}). Recall that the motivation is to facilitate integration with an SDR platform, the experimentation of FD at the link/network level, and integration with open-access wireless testbeds. 

\subsection{FDE PCB Canceller Implementation}
\label{ssec:impl-pcb}

\begin{figure}[!t]
\centering
\includegraphics[width=\columnwidth]{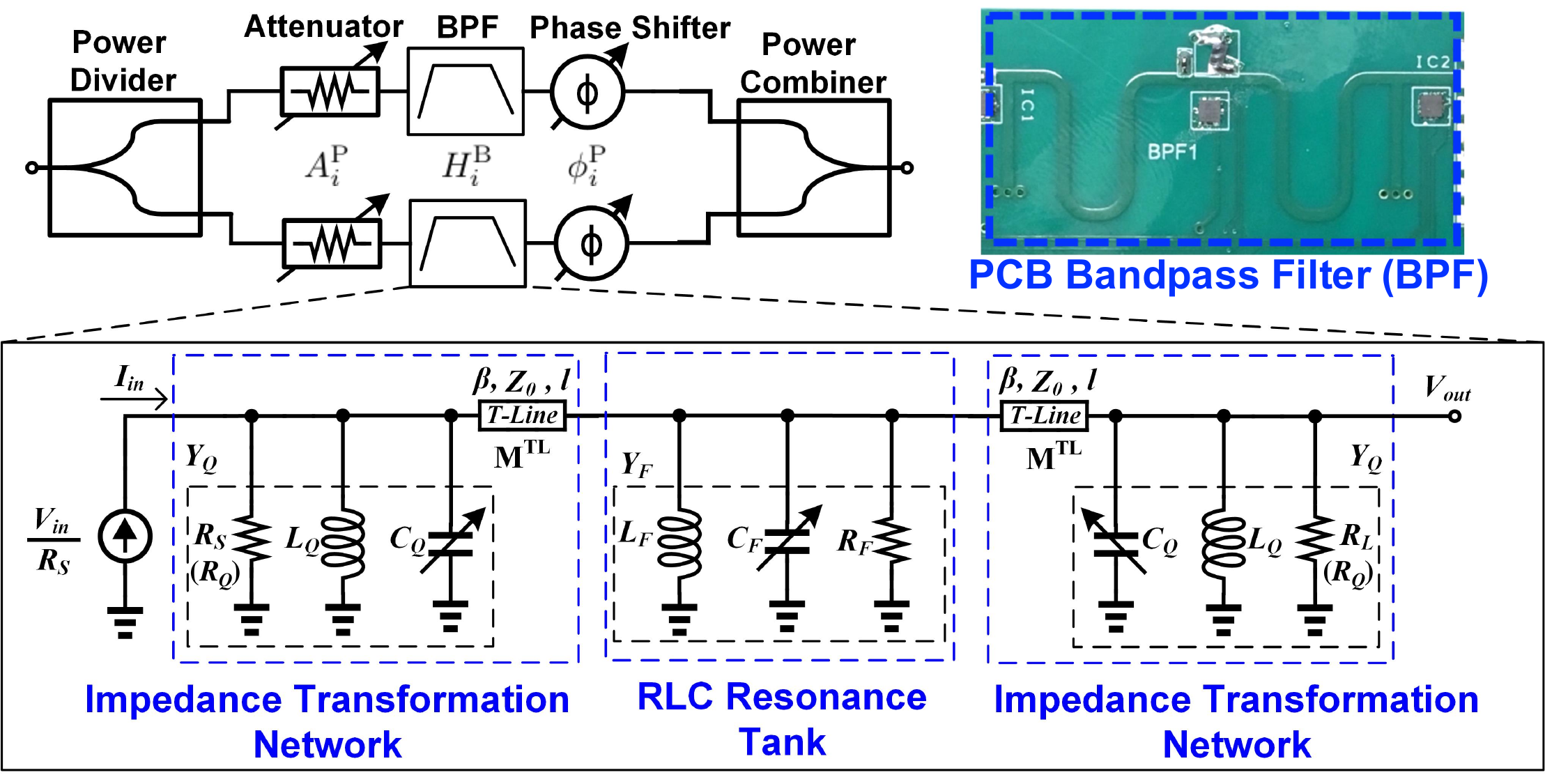}
\vspace{-\baselineskip}
\caption{Block diagram of the implemented 2 FDE taps in the PCB canceller (see Fig.~\ref{fig:fde-concept}(a)), each of which consists of an RLC bandpass filter (BPF), an attenuator for amplitude control, and a phase shifter for phase control.}
\label{fig:diagram-pcb}
\vspace{-0.5\baselineskip}
\end{figure}

Fig.~\ref{fig:intro}\subref{fig:intro-pcb} and Fig.~\ref{fig:fde-concept}\subref{fig:fde-concept-diagram} show the implementation and block diagram of the PCB canceller with \addedMK{two} FDE taps. \addedMK{A reference signal is coupled from the TX input and split to the two FDE taps through a power divider.} Then, the signals after each FDE tap are combined and RF SIC is performed at the RX input. Each FDE tap consists of a reconfigurable 2$^{\textrm{nd}}$-order BPF, \addedMK{and} an attenuator and phase shifter for amplitude and phase controls. We refer to the BPF here as the \emph{PCB BPF} to distinguish from the one in the RFIC canceller {\eqref{eq:rfic-tf}}. The PCB BPF (with size of {1.5}\thinspace{cm}$\times${4}\thinspace{cm}, see Fig.~\ref{fig:diagram-pcb}) is implemented as an RLC filter with impedance transformation networks and is optimized around {900}\thinspace{MHz} operating frequency.\footnote{We select {900}\thinspace{MHz} around the Region 2 {902--928}\thinspace{MHz} ISM band as the operating frequency but the approach can be easily extended to other bands (e.g., {2.4}\thinspace{GHz}) with slight modification of the hardware design and proper choice of the frequency-dependent components.} When compared to the $N$-path filter used in the RFIC canceller~\cite{Zhou_WBSIC_JSSC15} \addedMK{that has constant DC power draw, the PCB BPF} has zero DC power consumption. \addedMK{It can also support higher TX power levels and has a lower noise figure.}

The PCB BPF center frequency in the $i^{\textrm{th}}$ FDE tap can be adjusted through the capacitor, $\PCBTapCFCap{i}$, in the RLC resonance tank. In order to achieve a high and adjustable BPF quality factor, impedance transformation networks including transmission-lines (T-Lines) and digitally tunable capacitors, $\PCBTapQFCap{i}$, are introduced. In our implementation, $\PCBTapCFCap{i}$ consists of two parallel capacitors: a fixed {8.2}\thinspace{pF} capacitor and a digitally tunable capacitor (4-bit) with a {0.12}\thinspace{pF} resolution. For $\PCBTapQFCap{i}$, we use a digitally tunable capacitor (5-bit) with a {0.39}\thinspace{pF} resolution. In addition, the programmable attenuator has a tuning range of {0--15.5}\thinspace{dB} with a {0.5}\thinspace{dB} resolution, and the passive phase shifter is controlled by an 8-bit digital-to-analog converter (DAC) and covers full {360}$^{\circ}$ range.

\subsection{FDE PCB Canceller Model}
\label{ssec:impl-model}
Ideally, the PCB BPF has a 2$^{\textrm{nd}}$-order frequency response from the RLC resonance tank. However, in practical implementation, its response deviates from that used in the FDE-based RFIC canceller {\eqref{eq:rfic-tf}}. Therefore, there is a need for a valid model tailored for evaluating the performance and optimized configuration of the PCB canceller. Based on the circuit diagram in Fig.~\ref{fig:diagram-pcb}, we derive a realistic model for the frequency response of the PCB BPF, $\BPFTapTF{i}(f_k)$, given by\footnote{The details can be found in~\cite{chen2019wideband}.}
\begin{align}
\label{eq:pcb-bpf-tf}
\BPFTapTF{i}(f_k) = & R_s^{-1} \Big[ j\sin(2\beta l) Z_0Y_{\textrm{F},i}(f_k)Y_{\textrm{Q},i}(f_k) \nonumber \\
& + \cos^2(\beta l)Y_{\textrm{F},i}(f_k) + 2\cos(2\beta l)Y_{\textrm{Q},i}(f_k) \nonumber \\
& + j\sin(2\beta l)/Z_0 + 0.5j\sin(2\beta l)Z_0(Y_{\textrm{Q},i}(f_k))^2 \nonumber \\
& - \sin^2(\beta l)Z_0^2Y_{\textrm{F},i}(f_k)(Y_{\textrm{Q},i}(f_k))^2  \Big]^{-1},
\end{align}
where $Y_{\textrm{F},i}(f_k)$ and $Y_{\textrm{Q},i}(f_k)$ are the admittance of the RLC resonance tank and impedance transformation networks, i.e., 
\begin{equation}
\left\{
\begin{aligned}
\label{eq:pcb-admittance}
Y_{\textrm{F},i}(f_k) & = 1/R_{\textrm{F}} + j2\pi \PCBTapCFCap{i} f_k + 1/(j2\pi L_{\textrm{F}} f_k), \\
Y_{\textrm{Q},i}(f_k) & = 1/R_{\textrm{Q}} + j2\pi \PCBTapQFCap{i} f_k + 1/(j2\pi L_{\textrm{Q}} f_k).
\end{aligned}
\right.
\end{equation}
In particular, to have perfect matching with the source and load impedance of the RLC resonance tank, $R_{\textrm{S}}$ and $R_{\textrm{L}}$ are set to be the same value of $R_\textrm{Q} = {50}\thinspace\Omega$ (see Fig.~\ref{fig:diagram-pcb}). $\beta$ and $Z_0$ are the wavenumber and characteristic impedance of the T-Line with length $l$ (see Fig.~\ref{fig:diagram-pcb}). In our implementation, $L_{\textrm{F}} = 1.65\thinspace\textrm{nH}$, $L_{\textrm{Q}} = 2.85\thinspace\textrm{nH}$, $\beta l \approx 1.37\thinspace\textrm{rad}$, and $Z_0 = {50}\thinspace\Omega$.

In addition, other components in the PCB canceller (e.g., couplers and power divider/combiner) can introduce extra attenuation and group delay, due to implementation losses. Based on the S-Parameters of the components and measurements, we observed that the attenuation and group delay introduced, denoted by $\Amp{0}^{\textrm{P}}$ and $\tau_0^{\textrm{P}}$, are constant across frequency in the desired bandwidth. Hence, we empirically set $\Amp{0} = {-4.1}\thinspace\textrm{dB}$ and $\tau_0 = {4.2}\thinspace\textrm{ns}$. Recall that each FDE tap is also associated with amplitude and phase controls, $\PCBTapAmp{i}$ and $\PCBTapPhase{i}$, the PCB canceller with two FDE taps is modeled by
\begin{align}
\label{eq:pcb-tf-calibrated}
\PCBTF(f_k) & = \Amp{0}^{\textrm{P}} e^{-j2\pi f_k\tau_0^{\textrm{P}}} \left[ \littlesum_{i=1}^{2} \PCBTapAmp{i} e^{-j\PCBTapPhase{i}} \BPFTapTF{i}(f_k) \right],
\end{align}
where $\BPFTapTF{i}(f_k)$ is the PCB BPF model given by {\eqref{eq:pcb-bpf-tf}}. As a result, the $i^{\textrm{th}}$ FDE tap in the PCB canceller {\eqref{eq:pcb-tf-calibrated}} has configuration parameters $\{\PCBTapAmp{i}, \PCBTapPhase{i}, \PCBTapCFCap{i}, \PCBTapQFCap{i}\}$, featuring 4 DoF.

\subsection{Optimization of Canceller Configuration}
\label{ssec:impl-opt}
Based on {\OptProblem}, we now present a general FDE-based canceller configuration scheme that jointly optimizes all the FDE taps in the canceller.
Although our implemented PCB canceller has only 2 FDE taps, both its model and the configuration scheme can be easily extended to the case with a larger number of \addedMK{FDE taps~\cite{chen2019wideband}}.

The inputs to the FDE-based canceller configuration scheme are: (i) the PCB canceller model {\eqref{eq:pcb-tf-calibrated}} with given number of FDE taps, $\NumTap$, (ii) the antenna interface response, $\AntTF(f_k)$, and (iii) the desired RF SIC bandwidth, $f_k \in [f_1,f_\NumChnl]$. 
The optimized canceller configuration is obtained by solving {\OptProblemPCB}.
\begin{align*}
\OptProblemPCB\ & \min: \littlesum_{k=1}^{\NumChnl} \NormTwo{\PCBResTF(f_k)} = \littlesum_{k=1}^{\NumChnl} \NormTwo{ \AntTF(f_k) - \PCBTF(f_k) }^2 \\
\textrm{s.t.:}\ & \PCBTapAmp{i} \in [\PCBTapAmpMin, \PCBTapAmpMax],\ \PCBTapPhase{i} \in [-\pi, \pi], \\
& \PCBTapCFCap{i} \in [\PCBTapCFCapMin, \PCBTapCFCapMax],\ \PCBTapQFCap{i} \in [\PCBTapQFCapMin, \PCBTapQFCapMax],\ \forall i.
\end{align*}

Note that {\OptProblemPCB} is challenging to solve due to its non-convexity and non-linearity, as opposed to the linear program used in the delay line-based RF canceller~\cite{bharadia2013full}. This is due to the specific forms of the configuration parameters in {\eqref{eq:pcb-tf-calibrated}} such as (i) the higher-order terms introduced by $f_k$, and (ii) the trigonometric term introduced by the phase control, $\PCBTapPhase{i}$. In addition, the antenna interface response, $\AntTF(f_k)$, is also frequency-selective and time-varying.

In general, it is difficult to maintain analytical tractability of {\OptProblemPCB} (i.e., to obtain its optimal solution in closed-form). However, in practice, it is unnecessary to obtain the global optimum to {\OptProblemPCB} as long as the performance achieved by the obtained local optimum is sufficient (e.g., at least {45}]\thinspace{dB} RF SIC is achieved). In this work, the local optimal solution to (P2) is obtained using a MATLAB solver.
The detailed implementation and performance of the optimized canceller configuration are described in Section~\ref{ssec:exp-node}, and a numerical verification of the resulting improved SIC performance is presented in~\cite{chen2019wideband}.


%% file: tex/experiments_cosmos.tex
In this section, we discuss the integration of the PCB canceller described in Section~\ref{sec:impl} \addedMK{with the mobile FD testbed. Then, we present an extensive testbed evaluation of the mobile FD radios} at the node, link, and network levels.

\subsection{Implementation of the Mobile FD Testbed}
\label{ssec:exp-testbed}

\addedMK{The overall design of the mobile FD radios and SDR testbed is shown in Figs.~\ref{fig:intro}\subref{fig:intro-fd-radio} and~\ref{fig:intro}\subref{fig:intro-fd-net}. Each FD radio uses an 860-960\thinspace{MHz} circulator for the antenna interface, and the PCB canceller serves as the frontend of a USRP-2942 SDR with an SBX-120 daughterboard. Each of the three FD radios in the mobile testbed operates at 900\thinspace{MHz} carrier frequency, and further radios without FD frontends are used as HD users as experimentation requires. Further details on the implementation of the mobile FD testbed are provided in~\cite{chen2019wideband}.}

\addedMK{The mobile FD radios are supported by an OFDM PHY layer running at 20\thinspace{MHz} bandwidth implemented in NI LabVIEW on the host PC. This PC also runs the optimized PCB canceller configuration scheme. Briefly,} the canceller configuration scheme, explained in further detail in~\cite{chen2019wideband}, has the following steps:
\begin{enumerate}[leftmargin=*,topsep=0pt]
\item[1.]
Measure the real-time antenna interface response, $\AntTF(f_k)$, using a preamble of 2 OFDM symbols;
\item[2.]
Solve for an initial PCB canceller configuration using optimization {\OptProblemPCB} based on the measured $\AntTF(f_k)$ and the canceller model {\eqref{eq:pcb-tf-calibrated}} (see Section~\ref{ssec:impl-opt});
\item[3.]
Perform a finer-grained local search and record the optimal canceller configuration (usually around 10 iterations).
\end{enumerate}

\begin{figure}[!t]
\centering
\vspace{-\baselineskip}
\subfloat[]{
\label{fig:exp-usrp-algo-iq}
\includegraphics[width=0.47\columnwidth]{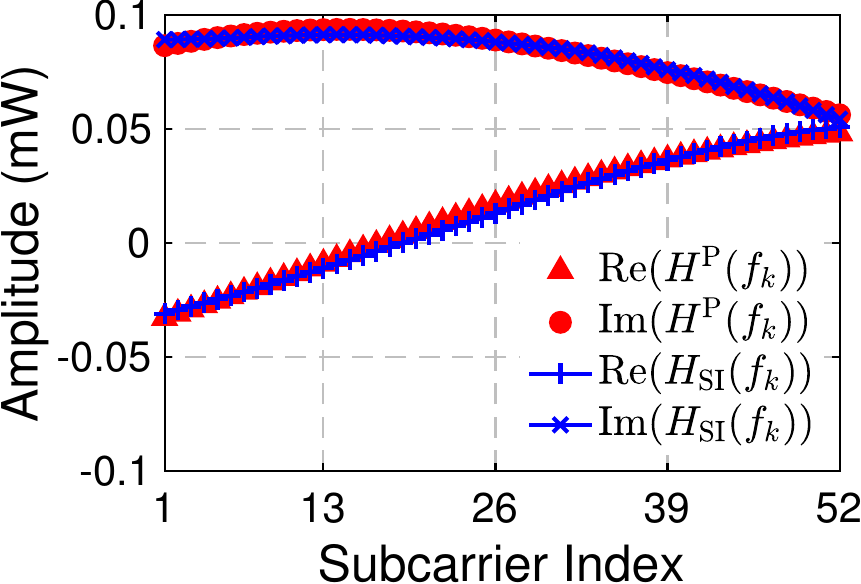}
}
\hfill
\subfloat[]{
\label{fig:exp-usrp-algo-res}
\includegraphics[width=0.47\columnwidth]{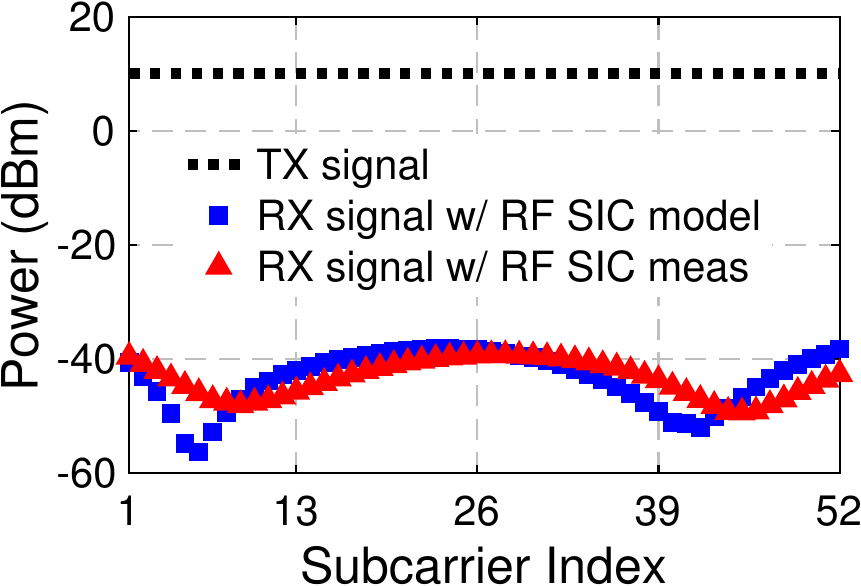}
}
\vspace{-0.5\baselineskip}
\caption{(a) Real and imaginary parts of the optimized PCB canceller response, $\PCBTF(f_k)$, vs. real-time SI channel measurements, $\AntTF(f_k)$, and (b) modeled and measured RX signal power after RF SIC at {10}\thinspace{dBm} TX power. An average {52}\thinspace{dB} RF SIC across {20}\thinspace{MHz} is achieved in the experiments.}
\label{fig:exp-usrp-algo}
\vspace{-0.5\baselineskip}
\end{figure}
\subsection{Node-Level: Microbenchmarks}
\label{ssec:exp-node}
\noindent\textbf{Optimized PCB canceller response and RF SIC}.
We set up an FDE-based FD radio running the optimized PCB canceller configuration scheme and record the canceller configuration, measured $\AntTF(f_k)$, and measured residual SI power after RF SIC. The recorded canceller configuration is then used to compute the PCB canceller response using {\eqref{eq:pcb-tf-calibrated}}.

Fig.~\ref{fig:exp-usrp-algo}\subref{fig:exp-usrp-algo-iq} shows an example of the optimized PCB canceller response, $\PCBTF(f_k)$, and the measured antenna interface response, $\AntTF(f_k)$, in real and imaginary parts (or I and Q). It can be seen that $\PCBTF(f_k)$ closely matches with $\AntTF(f_k)$ with maximum differences in amplitude and phase are only {0.5}\thinspace{dB} and {2.5}$^{\circ}$, respectively. This confirms the accuracy of the PCB canceller model and the performance of the optimized canceller configuration. Fig.~\ref{fig:exp-usrp-algo}\subref{fig:exp-usrp-algo-res} shows the modeled (computed by subtracting the modeled canceller response from the measured $\AntTF(f_k)$) and measured RX signal power after RF SIC at $+${10}\thinspace{dBm} TX power. The results show that the FDE-based FD radio achieves an average {52}\thinspace{dB} RF SIC across {20}\thinspace{MHz} bandwidth, from which {20}\thinspace{dB} is obtained from the antenna interface isolation. Similar performance is observed throughout the experimental evaluation. Note that the profile of RF SIC is similar to that presented in the sensitivity analysis of the FDE-based cancellers~\cite{chen2019wideband}.

\begin{figure}[!t]
\centering
\includegraphics[width=0.7\columnwidth]{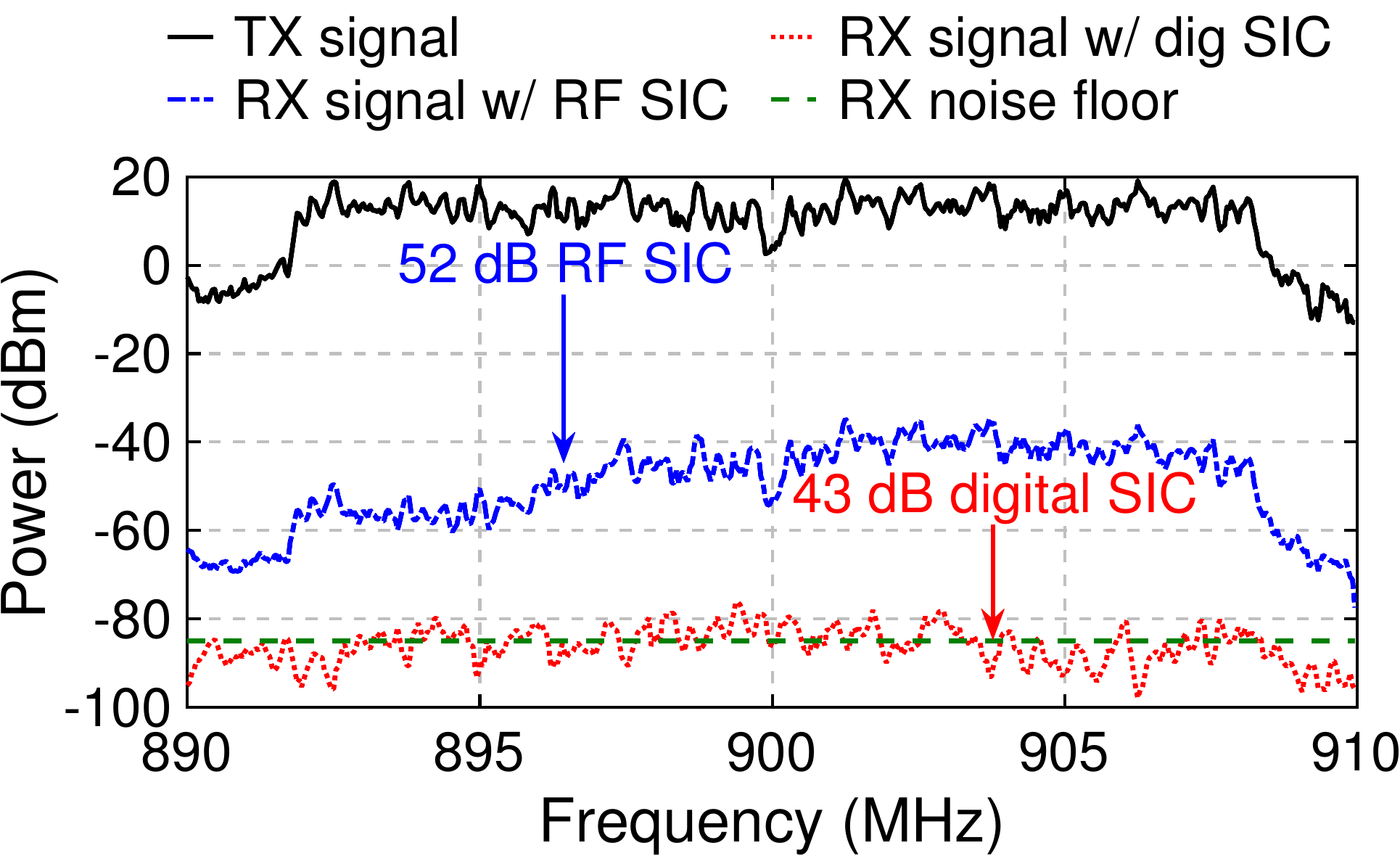}
\vspace{-0.5\baselineskip}
\caption{Power spectrum of the received signal after SIC in the RF and digital domains with {10}\thinspace{dBm} average TX power, {20}\thinspace{MHz} bandwidth, and $-${85}\thinspace{dBm} receiver noise floor.}
\label{fig:eval-usrp-spec-20mhz}
\vspace{-\baselineskip}
\end{figure}

\noindent\textbf{Overall SIC}.
We measure the overall SIC achieved by the FDE-based FD radio including the digital SIC in the same setting as described above, and the results are presented in Fig.~\ref{fig:eval-usrp-spec-20mhz}. It can be seen that the FDE-based FD radio achieves an average {95}\thinspace{dB} overall SIC across {20}\thinspace{MHz}, from which {52}\thinspace{dB} and {43}\thinspace{dB} are obtained in the RF and digital domains, respectively. Recall from Section~\ref{ssec:exp-testbed} that the USRP has noise floor of $-${85}\thinspace{dBm}, the FDE-based RF radio supports a maximal average TX power of {10}\thinspace{dBm} (where the peak TX power can go as high as {20}\thinspace{dBm}). We use TX power levels lower than or equal to {10}\thinspace{dBm} throughout the experiments, where all the SI can be canceled to below the RX noise floor.

\subsection{Link-Level: SNR Difference and FD Gain}
\label{ssec:exp-link}

\begin{figure}[!t]
\centering
\vspace{1\baselineskip}
\subfloat[LOS deployment and an FD radio in a hallway]{
\label{fig:exp-map-los}
\includegraphics[width=0.95\columnwidth]{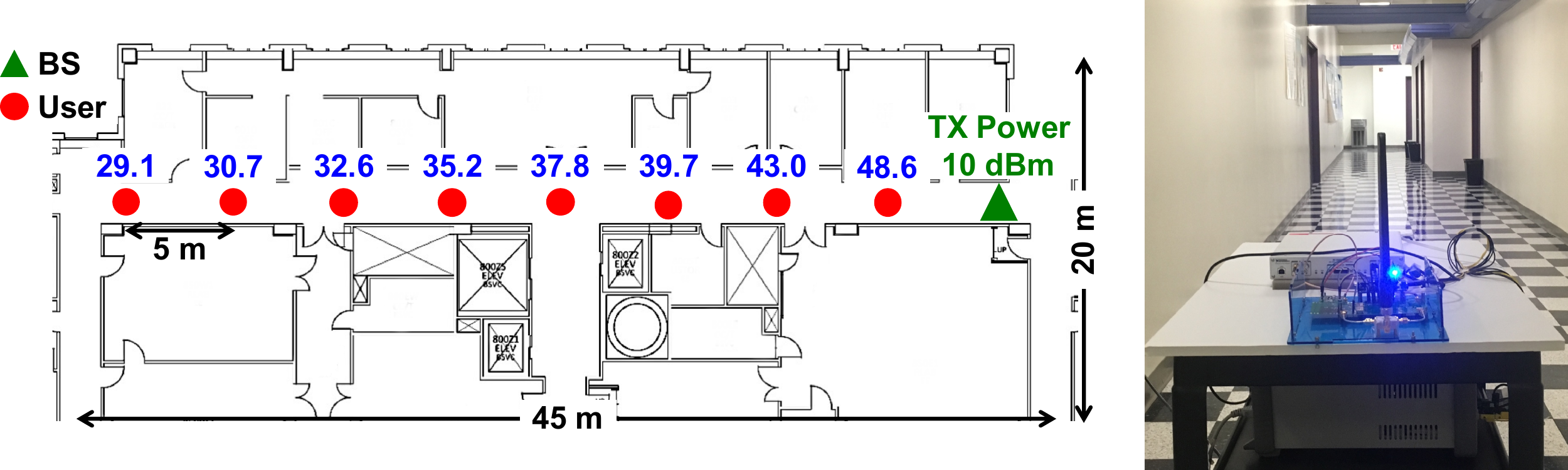}
} 
\\
\vspace{-0.5\baselineskip}
\subfloat[NLOS deployment and an FD radio in a lab environment]{
\label{fig:exp-map-nlos}
\includegraphics[width=0.95\columnwidth]{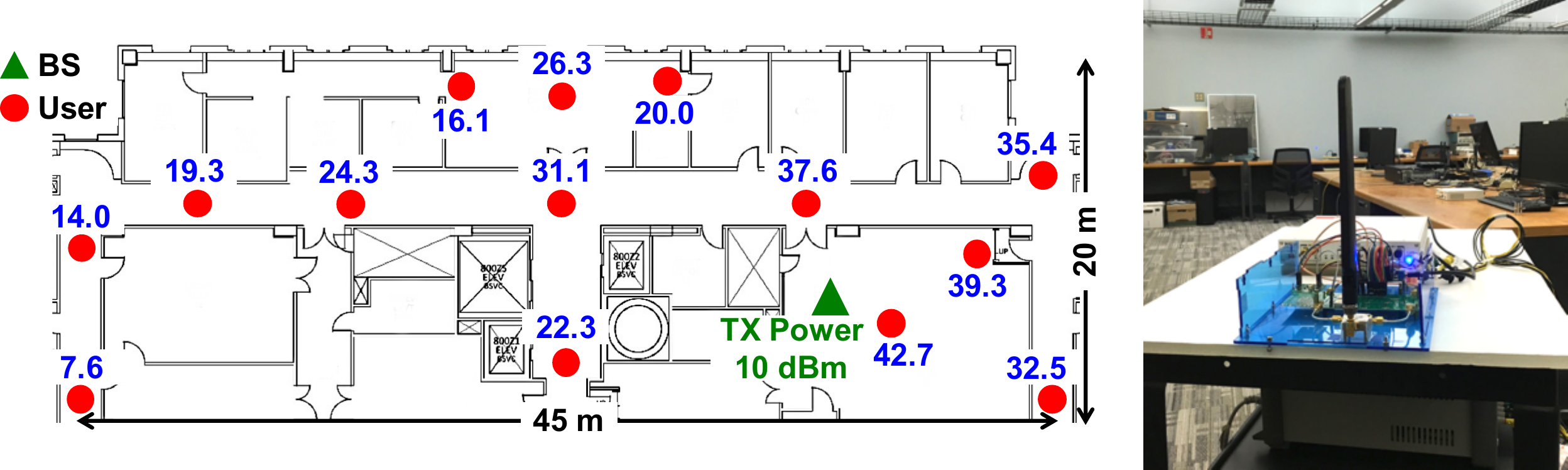}
}
\vspace{-0.5\baselineskip}
\caption{(a) Line-of-sight (LOS), and (b) non-line-of-sight (NLOS) deployments, and the measured HD link SNR values (dB).}
\label{fig:exp-map}
\vspace{-\baselineskip}
\end{figure}

\noindent\textbf{Experimental setup}.
To thoroughly evaluate the link level FD throughput gain achieved by our FD radio design, we conduct experiments with two FD radios with {10}\thinspace{dBm} TX power, one emulating a base station (BS) and one emulating a user. We consider both line-of-sight (LOS) and non-line-of-sight (NLOS) experiments as shown in Fig.~\ref{fig:exp-map}. In the LOS setting, the BS is placed at the end of a hallway and the user is moved away from the BS at stepsizes of 5 meters up to a distance of 40 meters. In the NLOS setting, the BS is placed in a lab environment with regular furniture and the user is placed at various locations (offices, labs, and corridors). We place the BS and the users at about the same height across all the experiments.\footnote{In this work, we emulate the BS and users without focusing on specific deployment scenarios. The impacts of different antenna heights and user densities, as mentioned in~\cite{lopez2015towards}, will be considered in future work.} The measured HD link SNR values are also included in Fig.~\ref{fig:exp-map}. Following the methodology of~\cite{bharadia2013full}, for each user location, we measure the \emph{link SNR difference}, which is defined as the absolute difference between the average HD and FD link SNR values. Throughout the experiments, link SNR values between {0--50}\thinspace{dB} are observed.

\begin{figure}[!t]
\centering
\subfloat[LOS Experiment]{
\label{fig:exp-link-snr-loss-los}
\includegraphics[width=0.47\columnwidth]{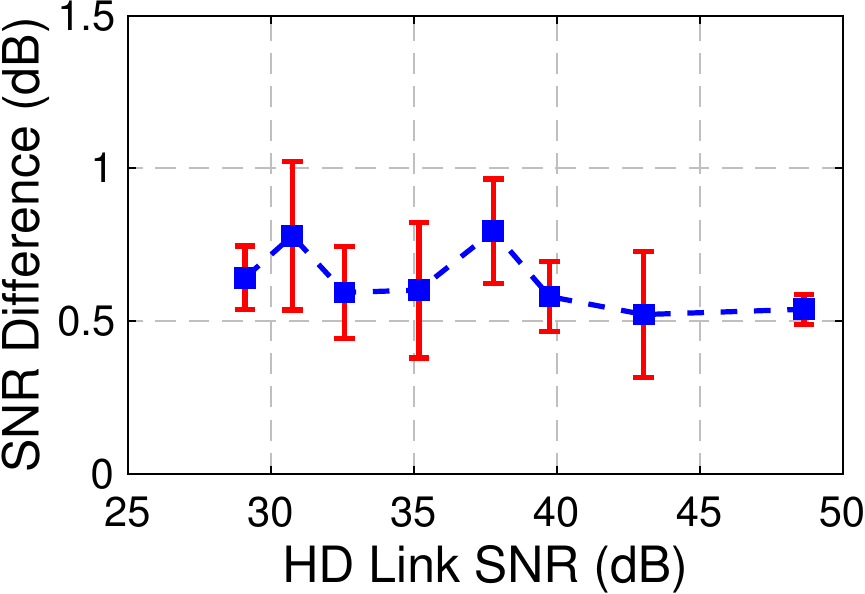}
}
\hfill
\subfloat[NLOS Experiment]{
\label{fig:exp-link-snr-loss-nlos}
\includegraphics[width=0.47\columnwidth]{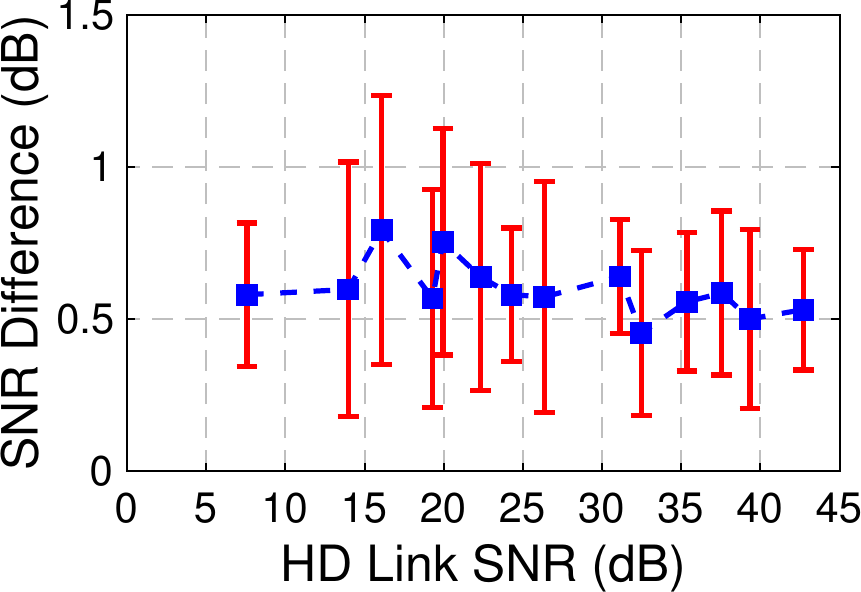}
}
\vspace{-0.5\baselineskip}
\caption{Difference between HD and FD link SNR values in the (a) LOS, and (b) NLOS experiments, with {10}\thinspace{dBm} TX power and 64QAM-3/4 MCS.}
\label{fig:exp-link-snr-loss}
\vspace{-0.5\baselineskip}
\end{figure}

\noindent\textbf{Difference in HD and FD link SNR values}.
Fig.~\ref{fig:exp-link-snr-loss} shows the measured link SNR difference as a function of the HD link SNR (i.e., for different user locations) in the LOS and NLOS experiments, respectively, with 64QAM-3/4 MCS. For the LOS experiments, the average link SNR difference is {0.6}\thinspace{dB} with a standard deviation of {0.2}\thinspace{dB}. For the NLOS experiments, the average link SNR difference is {0.6}\thinspace{dB} with a standard deviation of {0.3}\thinspace{dB}. The SNR difference has a higher variance in the NLOS experiments, due to the complicated environments (e.g., wooden desks and chairs, metal doors and bookshelves, etc.). In both cases, the link SNR difference is minimal and uncorrelated with user locations, showing the robustness of the FDE-based FD radio.

\begin{figure}[!t]
\centering
\vspace{-\baselineskip}
\subfloat[]{
\label{fig:exp-constellation-snr-loss}
\includegraphics[width=0.47\columnwidth]{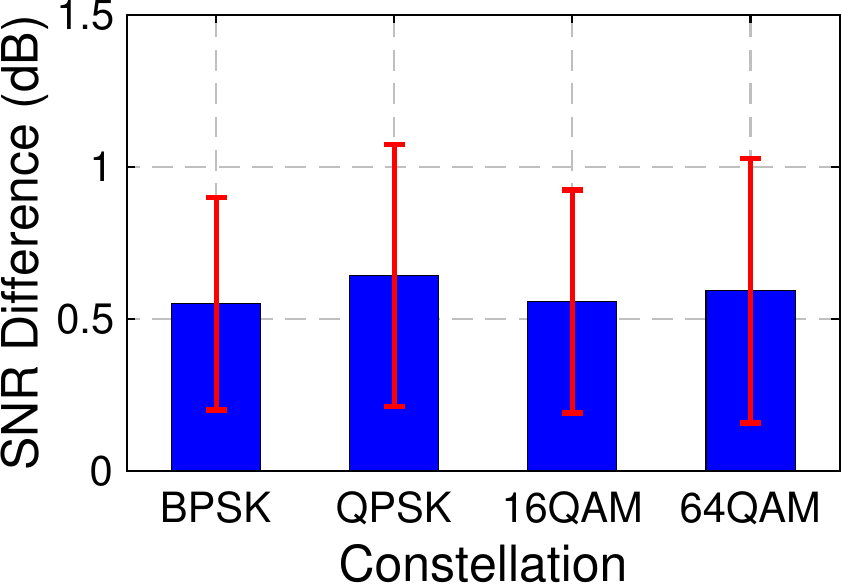}
}
\hfill
\subfloat[]{
\label{fig:exp-constellation-tput-gain}
\includegraphics[width=0.47\columnwidth]{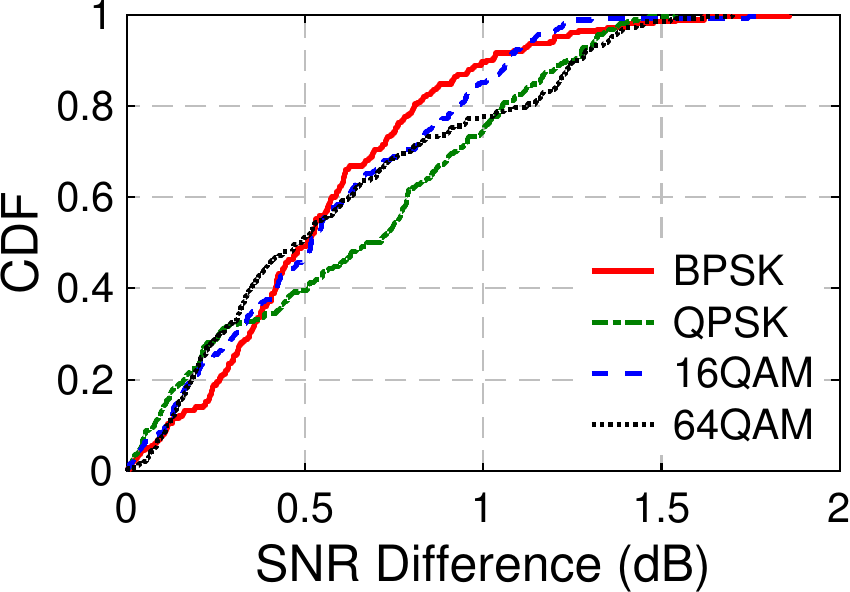}
}
\vspace{-0.5\baselineskip}
\caption{Difference between HD and FD link SNR values with {10}\thinspace{dBm} TX power under varying constellations: (a) mean and standard deviation, (b) CDF.}
\label{fig:exp-constellation}
\vspace{-0.5\baselineskip}
\end{figure}

\noindent\textbf{Impact of constellations}.
Fig.~\ref{fig:exp-constellation} shows the measured link SNR difference and its CDF with varying constellations and 3/4 coding rate. It can be seen that the link SNR difference has a mean of {0.6}\thinspace{dB} and a standard deviation of {0.4}\thinspace{dB}, both of which are uncorrelated with the constellations.

\noindent\textbf{FD link throughput and gain}.
For each user location in the LOS and NLOS experiments, the HD (resp.\ FD) link throughput is measured as the highest average data rate across all MCSs achieved by the link when both nodes operate in HD (resp.\ FD) mode . The FD gain is computed as the ratio between FD and HD throughput values. Recall that the maximal HD data rate is {54}\thinspace{Mbps}, an FD link data rate of {108}\thinspace{Mbps} can be achieved with an FD link PRR of 1.

\begin{figure}[!t]
\centering
\vspace{-\baselineskip}
\subfloat[LOS Experiment]{
\label{fig:exp-link-tput-los}
\includegraphics[width=0.47\columnwidth]{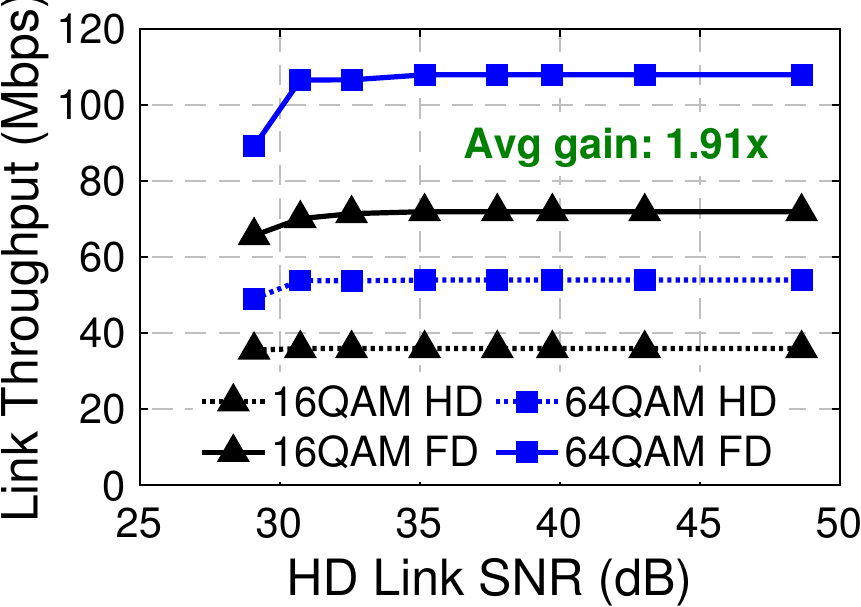}
}
\hfill
\subfloat[NLOS Experiment]{
\label{fig:exp-link-tput-nlos}
\includegraphics[width=0.47\columnwidth]{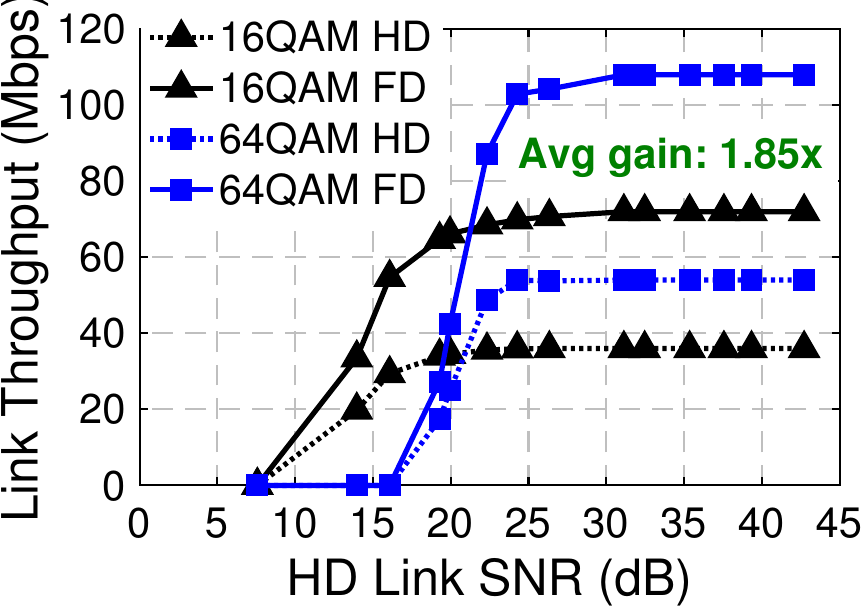}
}
\vspace{-0.5\baselineskip}
\caption{HD and FD link throughput in the (a) LOS, and (b) NLOS experiments, with {10}\thinspace{dBm} TX power and 16QAM-3/4 and 64QAM-3/4 MCSs.}
\label{fig:exp-link-tput}
\vspace{-0.5\baselineskip}
\end{figure}

Fig.~\ref{fig:exp-link-tput} shows the average HD and FD link throughput with varying MCSs, where each point represents the average throughput across 1,000 packets. The results show that with sufficient link SNR values (e.g., {30}\thinspace{dB} for 64QAM-3/4 MCS), the FDE-based FD radios achieve an \emph{exact} link throughput gain of 2$\times$. In these scenarios, the HD/FD link always achieves a link PRR of 1 which results in the maximum achievable HD/FD link data rate. With medium link SNR values, where the link PRR less than 1, the average FD link throughput gains across different MCSs are 1.91$\times$ and 1.85$\times$ for the LOS and NLOS experiments, respectively. \addedMK{We note that higher modulation schemes (e.g., 256QAM) will increase HD/FD throughput with high enough SNR, but FD gain is likely to remain similar. Such schemes are therefore not essential for evaluating the FDE-based canceller.}

\subsection{Network-Level FD Gain}
\label{ssec:exp-net}
We now experimentally evaluate the network-level throughput gain introduced by FD-capable BS and users. The users can significantly benefit from the FDE-based FD radio suitable for hand-held devices. We compare experimental results to the analysis (e.g.,~\cite{marasevic2017resource}) and demonstrate practical FD gain in different network settings. Specifically, we consider two types of networks as depicted in Fig.~\ref{fig:exp-net-setup}: (i) \emph{UL-DL networks} with one FD BS and two HD users with inter-user interference (IUI), and (ii) \emph{heterogeneous HD-FD networks} with HD and FD users. Due to software challenge with implementing a real-time MAC layer using a USRP, we apply a TDMA setting where each (HD or FD) user takes turn to be activated for the same period of time. \addedMK{In Section~\ref{sssec:cosmos-exp}, we describe a similar set of experiments conducted on the COSMOS FD testbed.}

\subsubsection{UL-DL Networks with IUI}
\label{sssec:exp-net-ul-dl}
\begin{figure}[!t]
\centering
\vspace{-\baselineskip}
\subfloat[]{
\label{fig:exp-net-setup-ul-dl}
\includegraphics[width=0.31\columnwidth]{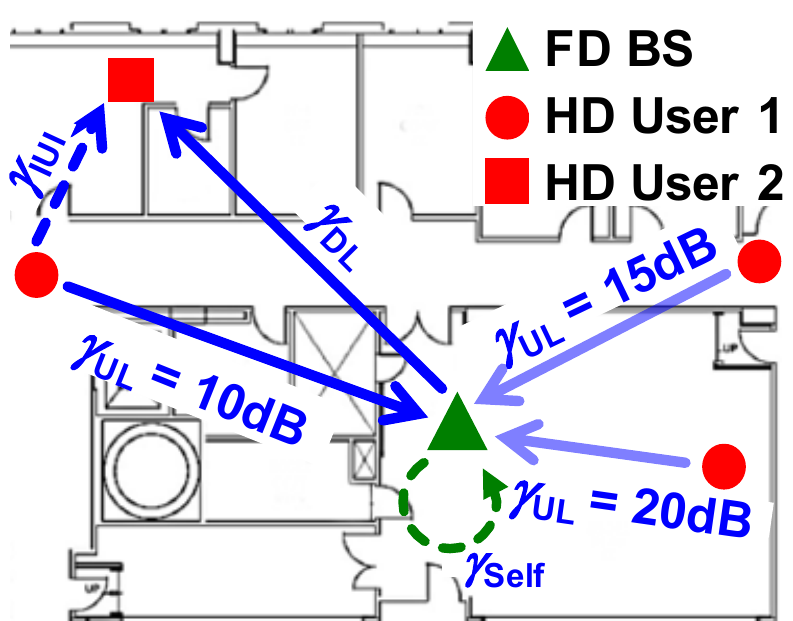}
}
\hspace{-6pt} \hfill
\subfloat[]{
\label{fig:exp-net-setup-two-users}
\includegraphics[width=0.31\columnwidth]{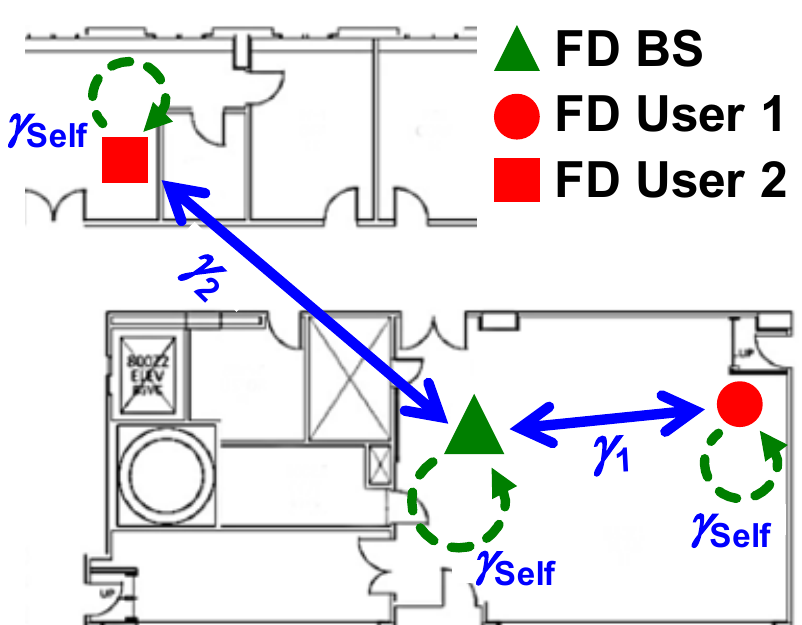}
}
\hspace{-6pt} \hfill
\subfloat[]{
\label{fig:exp-net-setup-three-users}
\includegraphics[width=0.31\columnwidth]{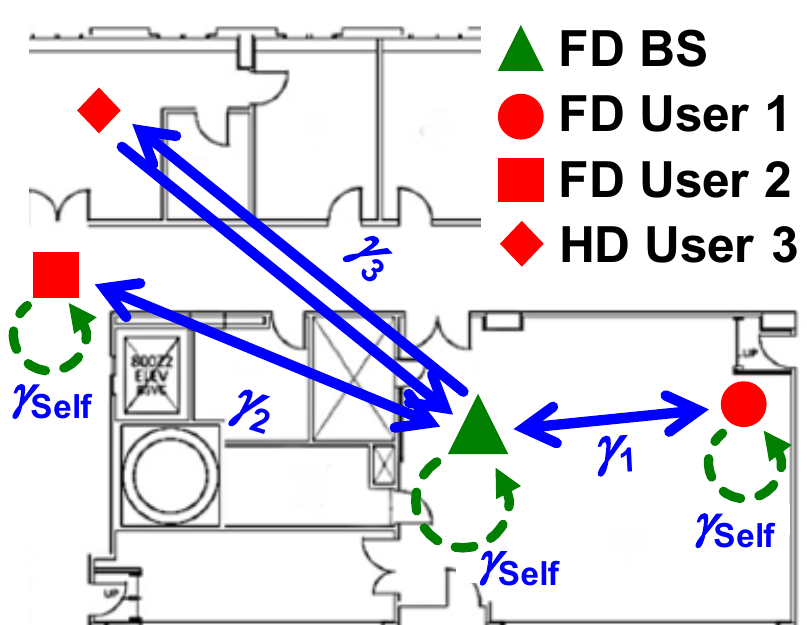}
}
\vspace{-0.5\baselineskip}
\caption{An example experimental setup for: (a) the UL-DL networks with varying $\SNRUL$ and $\SNRDL$, (b) heterogeneous 3-node network with one FD BS and 2 FD users, and (c) heterogeneous 4-node networks with one FD BS, 2 FD users, and one HD user.}
\label{fig:exp-net-setup}
\end{figure}

We first consider UL-DL networks consisting of one FD BS and two HD users (indexed 1 and 2). Without loss of generality, in this setting, user 1 transmits on the UL to the BS, and the BS transmits to user 2 on the DL (see Fig.~\ref{fig:exp-net-setup}\subref{fig:exp-net-setup-ul-dl}).

\noindent\textbf{Analytical FD gain}.
We use Shannon's capacity formula $\DataRate{}(\SNR{}) = \BW \cdot \log_{2}(1+\SNR{})$ to compute the \emph{analytical throughput} of a link under bandwidth $\BW$ and (HD) link SNR $\SNR{}$. If the BS is only HD-capable, the network throughput in a UL-DL network when the UL and DL share the channel in a TDMA manner with equal fraction of time is given by
\begin{align}
\label{eq:net-ul-dl-tput-hd}
\DataRate{\textrm{UL-DL}}^{\textrm{HD}} = \frac{\BW}{2}\cdot\log_{2}\left(1+\SNRUL\right) + \frac{\BW}{2}\cdot\log_{2}\left(1+\SNRDL\right),
\end{align}
where $\SNRUL$ and $\SNRDL$ are the UL and DL SNRs, respectively. If the BS is FD-capable, the UL and DL can be simultaneously activated with an analytical network throughput of
\begin{align}
\label{eq:net-ul-dl-tput-fd}
\DataRate{\textrm{UL-DL}}^{\textrm{FD}} = \BW\cdot\log_{2}(1+\frac{\SNRUL}{1+\XINR}) + \BW\cdot\log_{2}(1+\frac{\SNRDL}{1+\IUI}),
\end{align}
where: (i) $(\frac{\SNRDL}{1+\IUI})$ is the signal-to-interference-plus-noise ratio (SINR) at the DL HD user, and (ii) $\XINR$ is the residual self-interference-to-noise ratio (XINR) at the FD BS. We set $\XINR=1$ when computing the analytical throughput. Namely, the residual SI power is no higher than the RX noise floor (which can be achieved by the FDE-based FD radio, see Section~\ref{ssec:exp-node}). The \emph{analytical FD gain} is then defined as the ratio $(\DataRate{\textrm{UL-DL}}^{\textrm{FD}}/\DataRate{\textrm{UL-DL}}^{\textrm{HD}})$. Note that the FD gain depends on the coupling between $\SNRUL$, $\SNRDL$, and $\IUI$, which depend on the BS/user locations, their TX power levels, etc. 

\noindent\textbf{Experimental FD gain}.
The experimental setup is depicted in Fig.~\ref{fig:exp-net-setup}\subref{fig:exp-net-setup-ul-dl}, where the TX power levels of the BS and user 1 are set to be {10}\thinspace{dBm} and $-${10}\thinspace{dBm}, respectively. We fix the location of the BS and consider different UL SNR values of $\SNRUL = 10/15/20\thinspace\textrm{dB}$ by placing user 1 at three different locations. For each value of $\SNRUL$, user 2 is placed at 10 different locations, resulting in varying $\SNRDL$ and $\IUI$ values.

\begin{figure}[!t]
\centering
\vspace{-\baselineskip}
\subfloat[$\SNRUL={10}\thinspace\textrm{dB}$]{
\label{fig::exp-net-ul-dl-gain-10db}
\includegraphics[height=1.35in]{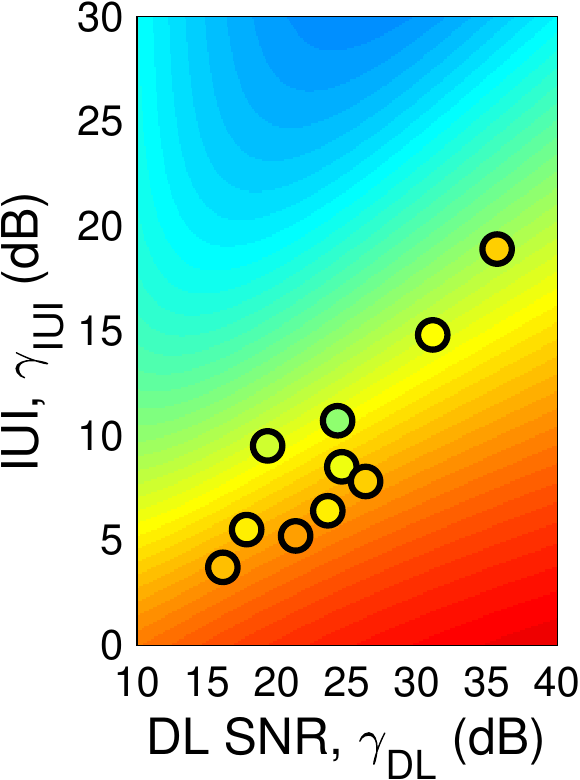}
}
\subfloat[$\SNRUL={15}\thinspace\textrm{dB}$]{
\label{fig::exp-net-ul-dl-gain-15db}
\includegraphics[height=1.35in]{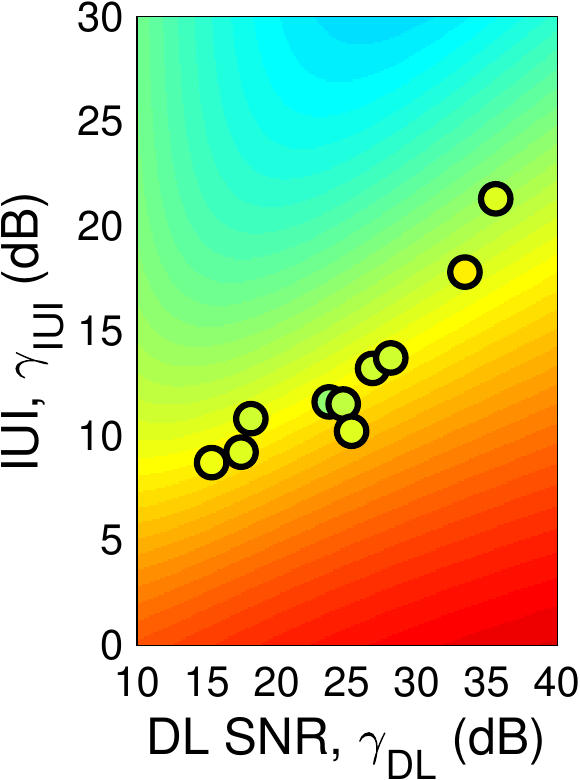}
}
\subfloat[$\SNRUL={20}\thinspace\textrm{dB}$]{
\label{fig::exp-net-ul-dl-gain-20db}
\includegraphics[height=1.35in]{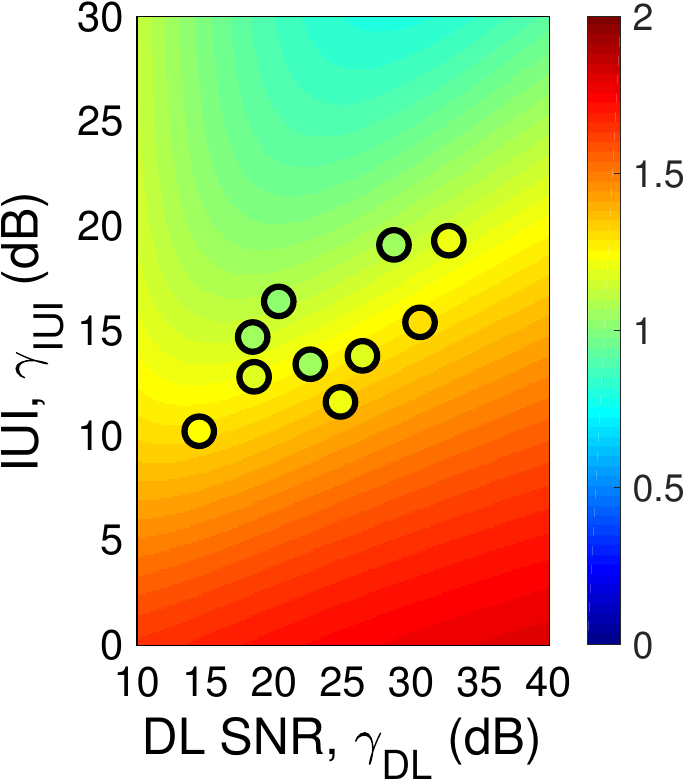}
}
\caption{Analytical (colored surface) and experimental (filled circles) network throughput gain for UL-DL networks consisting of one FD BS and two HD users with varying UL and DL SNR values, and inter-user interference (IUI) levels: (a) $\SNRUL={10}\thinspace\textrm{dB}$, (b) $\SNRUL={15}\thinspace\textrm{dB}$, and (c) $\SNRUL={20}\thinspace\textrm{dB}$. The baseline is the network throughput when the BS is HD.}
\label{fig:exp-net-ul-dl-gain}
\vspace{-0.5\baselineskip}
\end{figure}

Fig.~\ref{fig:exp-net-ul-dl-gain} shows the analytical (colored surface) and experimental (filled circles) FD gain, where the analytical gain is extracted using {\eqref{eq:net-ul-dl-tput-hd}} and {\eqref{eq:net-ul-dl-tput-fd}}, and the experimental gain is computed using the measured UL and DL throughput. It can be seen that smaller values of $\SNRUL$ and lower ratios between $\SNRDL$ and $\IUI$ lead to higher throughput gains in both analysis and experiments. The average analytical and experimental FD gains are summarized in Table~\ref{table:exp-net-ul-dl-gain}. Due to practical reasons such as the link SNR difference and its impact on link PRR~\cite{chen2019wideband}, the experimental FD gain is within 93\% of the analytical FD gain. The results confirm the analysis in~\cite{marasevic2017resource} and demonstrate the practical FD gain achieved in wideband UL-DL networks without any changes in the current network stack (i.e., only bringing FD capability to the BS). Moreover, performance improvements are expected through advanced power control and scheduling schemes.


\subsubsection{Heterogeneous 3-Node Networks}
\label{sssec:exp-net-two-users}
We consider heterogeneous HD-FD networks with 3 nodes: one FD BS and two users that can operate in either HD or FD mode (see an example experimental setup in Figs.~\ref{fig:intro}\subref{fig:intro-fd-net} and~\ref{fig:exp-net-setup}\subref{fig:exp-net-setup-two-users}). All 3 nodes have the same {0}\thinspace{dBm} TX power so that each user has symmetric UL and DL SNR values of $\SNR{i}$ ($i=1,2$).
We place user 1 at 5 different locations and place user 2 at 10 different locations for each location of user 1, resulting in a total number of 50 pairs of $(\SNR{1},\SNR{2})$. \addedMK{We also consider a similar experiment running on the COSMOS testbed in Section~\ref{sssec:cosmos-exp-tdma-3node}.}

\noindent\textbf{Analytical FD gain}.
We set the users to share the channel in a TDMA manner. The analytical network throughput in a 3-node network when zero, one, and two users are FD-capable is respectively given by
\begin{align}
& \hspace*{-10pt}
\DataRate{}^{\textrm{HD}} = \frac{\BW}{2}\cdot\log_{2}\left(1+\SNR{1}\right) + \frac{\BW}{2}\cdot\log_{2}\left(1+\SNR{2}\right), \label{eq:net-two-users-tput-hd} \\
& \hspace*{-10pt}
\DataRate{\textrm{User}~i~\textrm{FD}}^{\textrm{HD-FD}} = \BW\cdot\log_{2}(1+\frac{\SNR{i}}{1+\XINR}) + \frac{\BW}{2}\cdot\log_{2}(1+\SNR{\overline{i}}),\ \label{eq:net-two-users-tput-hd-fd} \\
& \hspace*{-10pt}
\DataRate{}^{\textrm{FD}} = \BW\log_{2}\cdot(1+\frac{\SNR{1}}{1+\XINR}) + \BW\log_{2}\cdot(1+\frac{\SNR{2}}{1+\XINR}), \label{eq:net-two-users-tput-fd}
\end{align}
where $\XINR=1$ is set (similar to Section~\ref{sssec:exp-net-ul-dl}). We consider both FD gains of $(\DataRate{\textrm{User}~i~\textrm{FD}}^{\textrm{HD-FD}}/\DataRate{}^{\textrm{HD}})$ (i.e., user $i$ is FD and user $\overline{i} \ne i$ is HD), and $(\DataRate{}^{\textrm{FD}}/\DataRate{}^{\textrm{HD}})$ (i.e., both users are FD).

\begin{figure}[!t]
\centering
\vspace{-\baselineskip}
\subfloat[Only user 1 FD]{
\label{fig:exp-net-two-users-user1-fd}
\includegraphics[height=1.35in]{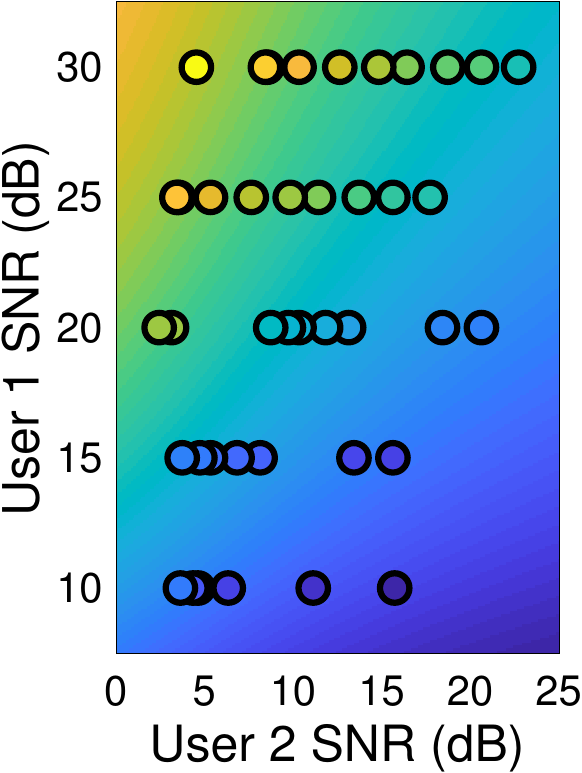}
}
\subfloat[Only user 2 FD]{
\label{fig:exp-net-two-users-user2-fd}
\includegraphics[height=1.35in]{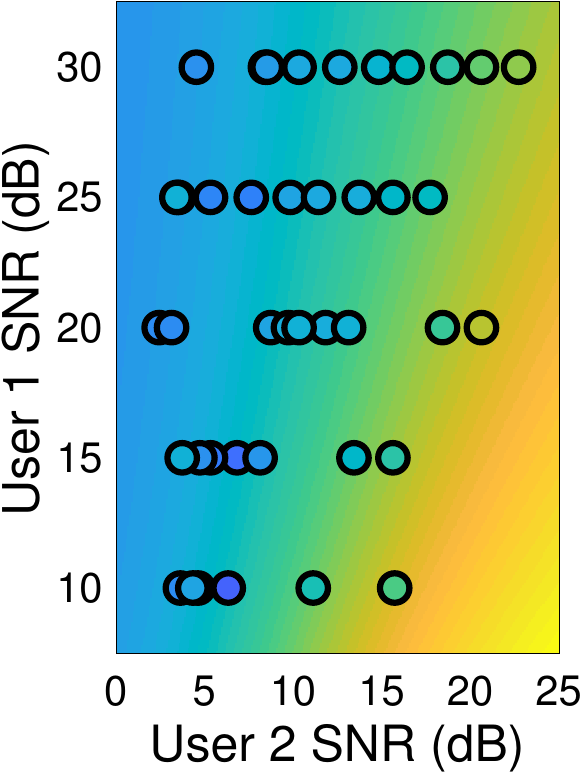}
}
\subfloat[Both users FD]{
\label{fig:exp-net-two-users-both-fd}
\includegraphics[height=1.35in]{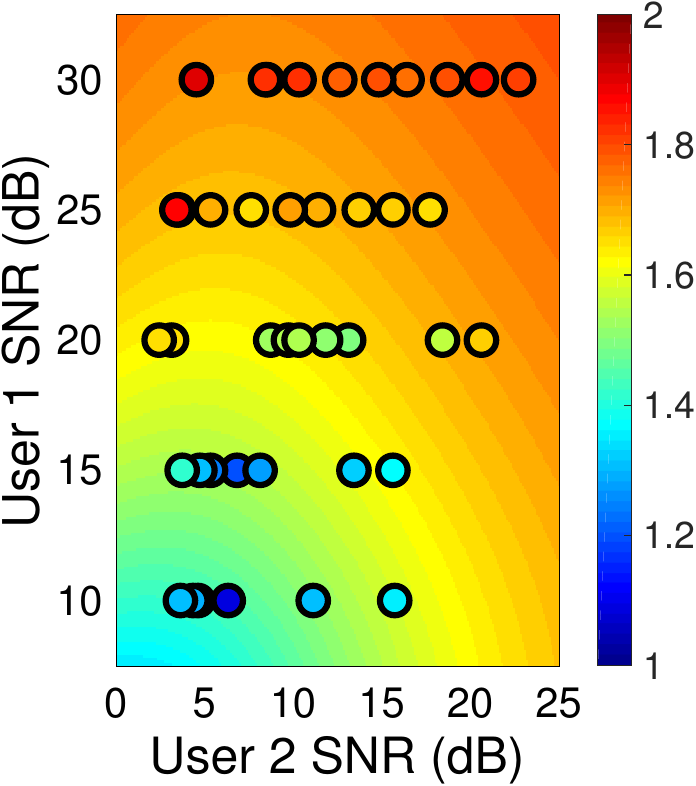}
}
\vspace{-0.5\baselineskip}
\caption{Analytical (colored surface) and experimental (filled circles) network throughput gain for 3-node networks consisting of one FD BS and two users with varying link SNR values: (a) only user 1 is FD, (b) only user 2 is FD, and (c) both users are FD. The baseline is the network throughput when both users are HD.}
\label{fig:exp-net-two-users}
\vspace{-0.5\baselineskip}
\end{figure}

\noindent\textbf{Experimental FD gain}.
For each pair of $(\SNR{1},\SNR{2})$, experimental FD gain is measured in three cases: (i) only user 1 is FD, (ii) only user 2 is FD, and (iii) both users are FD. Fig.~\ref{fig:exp-net-two-users} shows the analytical (colored surface) and experimental (filled circles) FD gain for each case. We exclude the results with $\SNR{i}<{3}\thinspace\textrm{dB}$ since the packets cannot be decoded, resulting in a throughput of zero.


The results show that with small link SNR values, the experimental FD gain is lower than the analytical value due to the inability to decode the packets. On the other hand, with sufficient link SNR values, the experimental FD gain exceeds the analytical FD gain. This is because setting $\XINR=1$ in {\eqref{eq:net-two-users-tput-hd-fd}} and {\eqref{eq:net-two-users-tput-fd}} results in a {3}\thinspace{dB} SNR loss in the analytical FD link SNR, and thereby in a lower throughput. However, in practice, the packets can be decoded with a link PRR of 1 with sufficient link SNRs, resulting in exact twice number of packets being successfully sent over an FD link. Moreover, the FD gain is more significant when enabling FD capability for users with higher link SNR values.

\begin{table}[!t]
\caption{Average FD Gain in UL-DL Networks with IUI.}
\label{table:exp-net-ul-dl-gain}
\vspace{-\baselineskip}
\footnotesize
\begin{center}
\begin{tabular}{|c|c|c|}
\hline
UL SNR, $\SNRUL$ & Analytical FD Gain & Experimental FD Gain \\
\hline
{10}\thinspace{dB} & 1.30$\times$ & 1.25$\times$ \\
\hline
{15}\thinspace{dB} & 1.23$\times$ & 1.16$\times$ \\
\hline
{20}\thinspace{dB} & 1.22$\times$ & 1.14$\times$ \\
\hline
\end{tabular}
\end{center}
\vspace{-2\baselineskip}
\end{table}

\begin{figure}[!t]
\centering
\subfloat[User 1 $\SNR{1}={15}\thinspace\textrm{dB}$]{
\label{fig:exp-net-two-users-fairness-snr1-15db}
\includegraphics[height=1.5in]{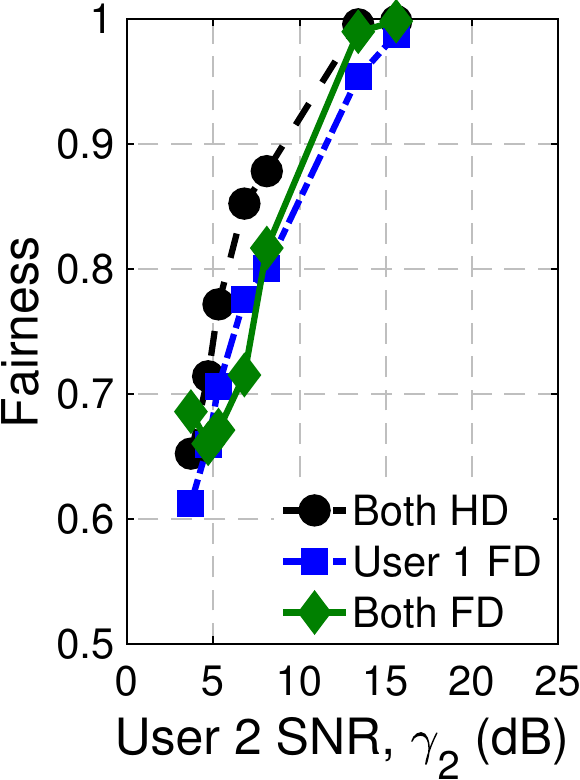}
}
\hspace{-10pt}
\subfloat[User 1 $\SNR{1}={20}\thinspace\textrm{dB}$]{
\label{fig:exp-net-two-users-fairness-snr1-20db}
\includegraphics[height=1.5in]{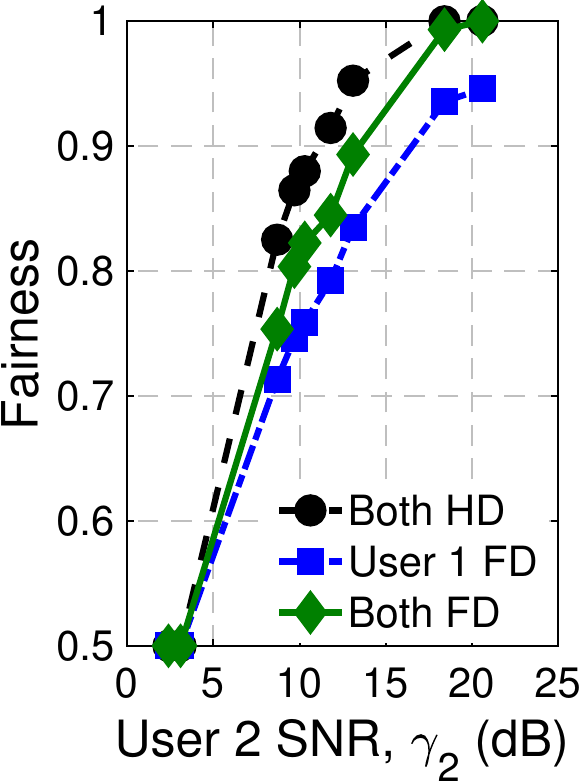}
}
\hspace{-10pt}
\subfloat[User 1 $\SNR{1}={25}\thinspace\textrm{dB}$]{
\label{fig:exp-net-two-users-fairness-snr1-25db}
\includegraphics[height=1.5in]{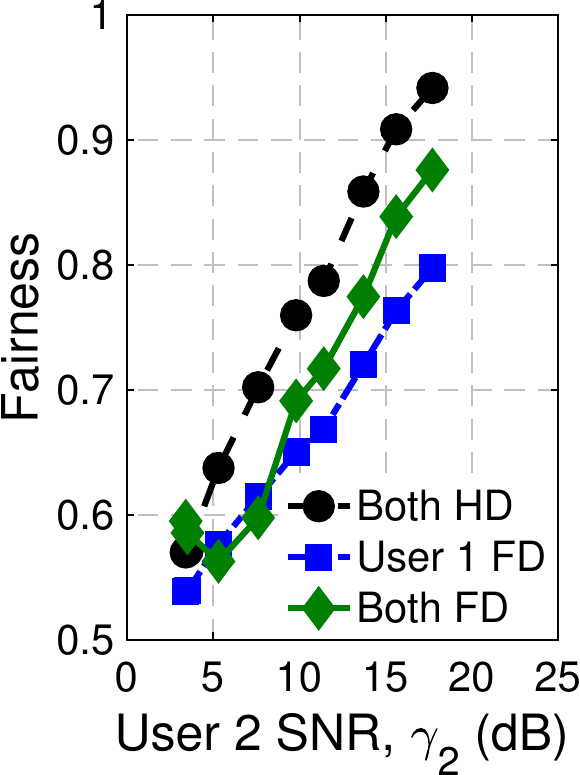}
}
\vspace{-0.5\baselineskip}
\caption{Measured Jain's fairness index (JFI) in 3-node networks where both users are HD, User 1 is FD, and both users are FD, with varying $(\SNR{1},\SNR{2})$.}
\label{fig:exp-net-two-users-fairness}
\vspace{-0.5\baselineskip}
\end{figure}

Another important metric we consider is the fairness between users, which is measured by the Jain's fairness index (JFI). In the considered 3-node networks, the JFI ranges between 0.5 (worst case) and 1 (best case). Fig.~\ref{fig:exp-net-two-users-fairness} shows the measured JFI when both users operate in HD mode, user 1 operates in FD mode, and both users operate in FD mode. with varying user SNR values $(\SNR{1},\SNR{2})$.
The results show that introducing FD capability to both users results in an average degradation in the network JFI of only 5.6/4.4/7.4\% for $\SNR{1} = {15/20/25}\thinspace\textrm{dB}$ (averaged across varying $\SNR{2}$), while the average network FD gains are 1.32/1.58/1.73$\times$ (see Fig.~\ref{fig:exp-net-two-users}), respectively. In addition, the JFI increases with more balanced user SNR values, which is as expected. For example, under the same value of $\SNR{1}$, increased value of $\SNR{2}$ (with $\SNR{2}<\SNR{1}$) leads to improved JFI, whose value approaches to 1 as $\SNR{2}$ approaches $\SNR{1}$.

\subsubsection{Heterogeneous 4-Node Networks}
\label{sssec:exp-net-three-users}

We experimentally study 4-node networks consisting of an FD BS and three users with {10}\thinspace{dBm}d TX power (see an example experimental setup in Fig.~\ref{fig:exp-net-setup}\subref{fig:exp-net-setup-three-users}).
The experimental setup is similar to that described in Section~\ref{sssec:exp-net-two-users}. 100 experiments are conducts where the 3 users are placed at different locations with different user SNR values. For each experiment, the network throughput is measured in three cases where: (i) zero, (ii) one, and (iii) two users are FD-capable.

\begin{figure}[!t]
\centering
\includegraphics[width=0.85\columnwidth]{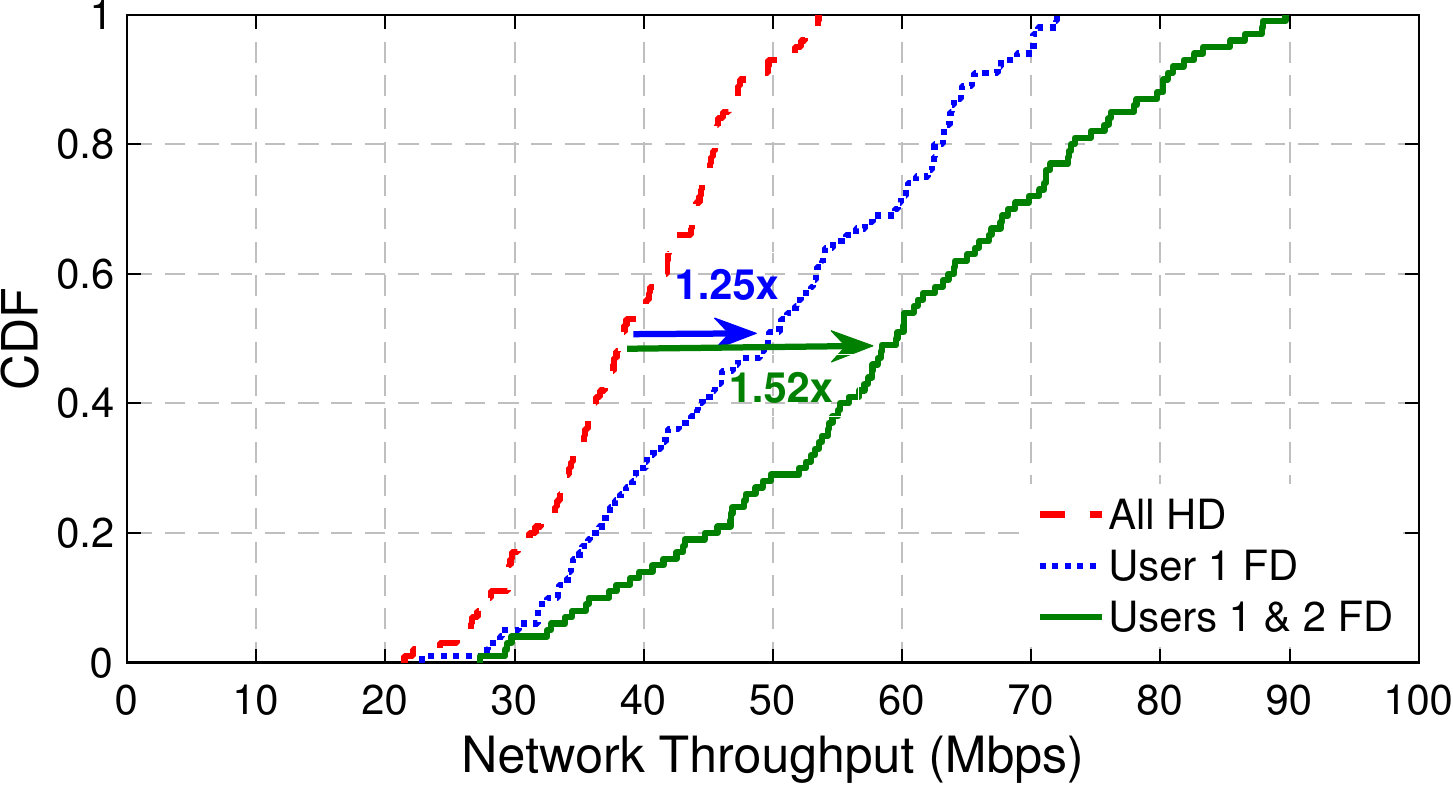}
\vspace{-0.5\baselineskip}
\caption{CDF of the experimental network throughput in 4-node networks when zero, one, or two users are FD-capable.}
\label{fig:exp-net-three-users}
\vspace{-\baselineskip}
\end{figure}

\begin{figure}[!t]
\centering
\includegraphics[width=0.85\columnwidth]{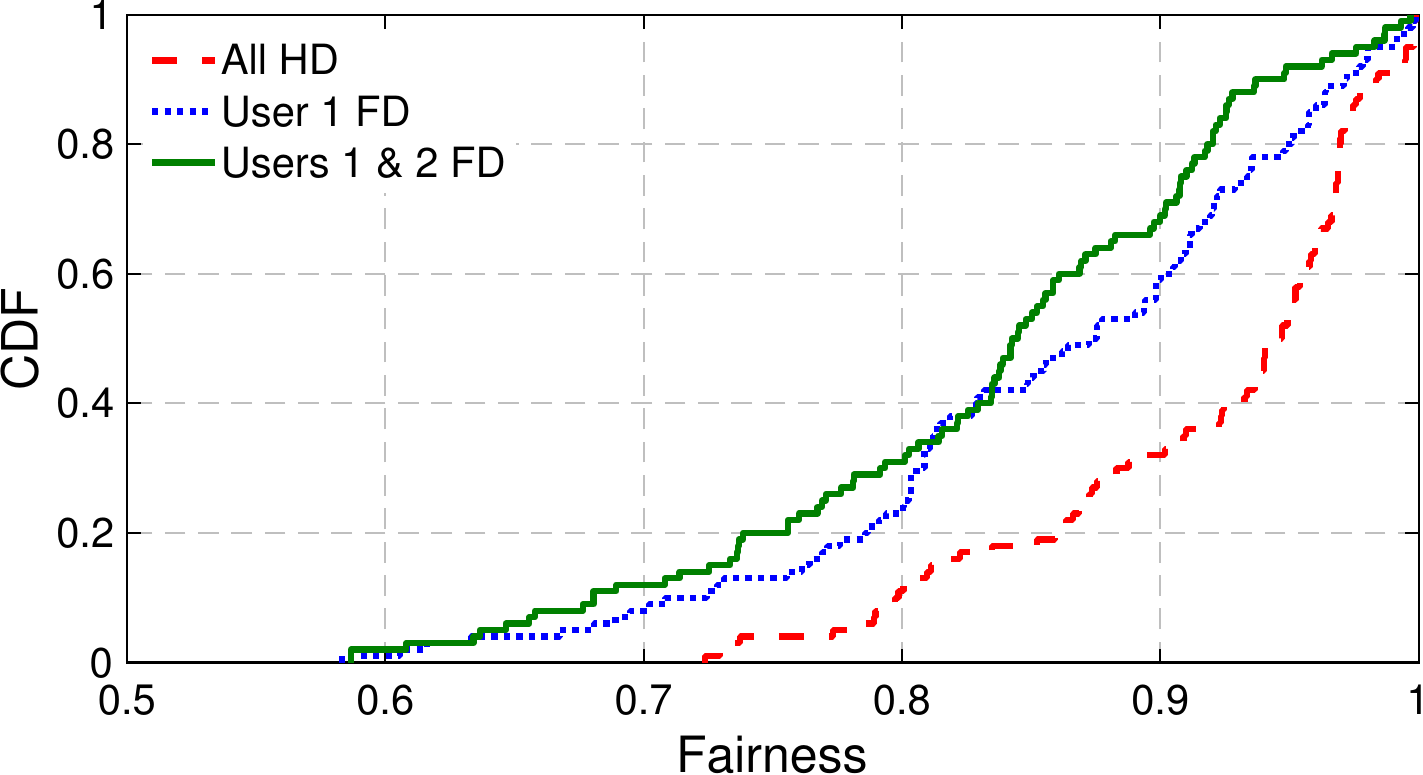}
\vspace{-0.5\baselineskip}
\caption{CDF of the measured Jain's fairness index (JFI) in 4-node networks when zero, one, or two users are FD-capable.}
\label{fig:exp-net-three-users-fairness}
\vspace{-\baselineskip}
\end{figure}

\begin{figure}[!t]
\centering
\includegraphics[width=0.85\columnwidth]{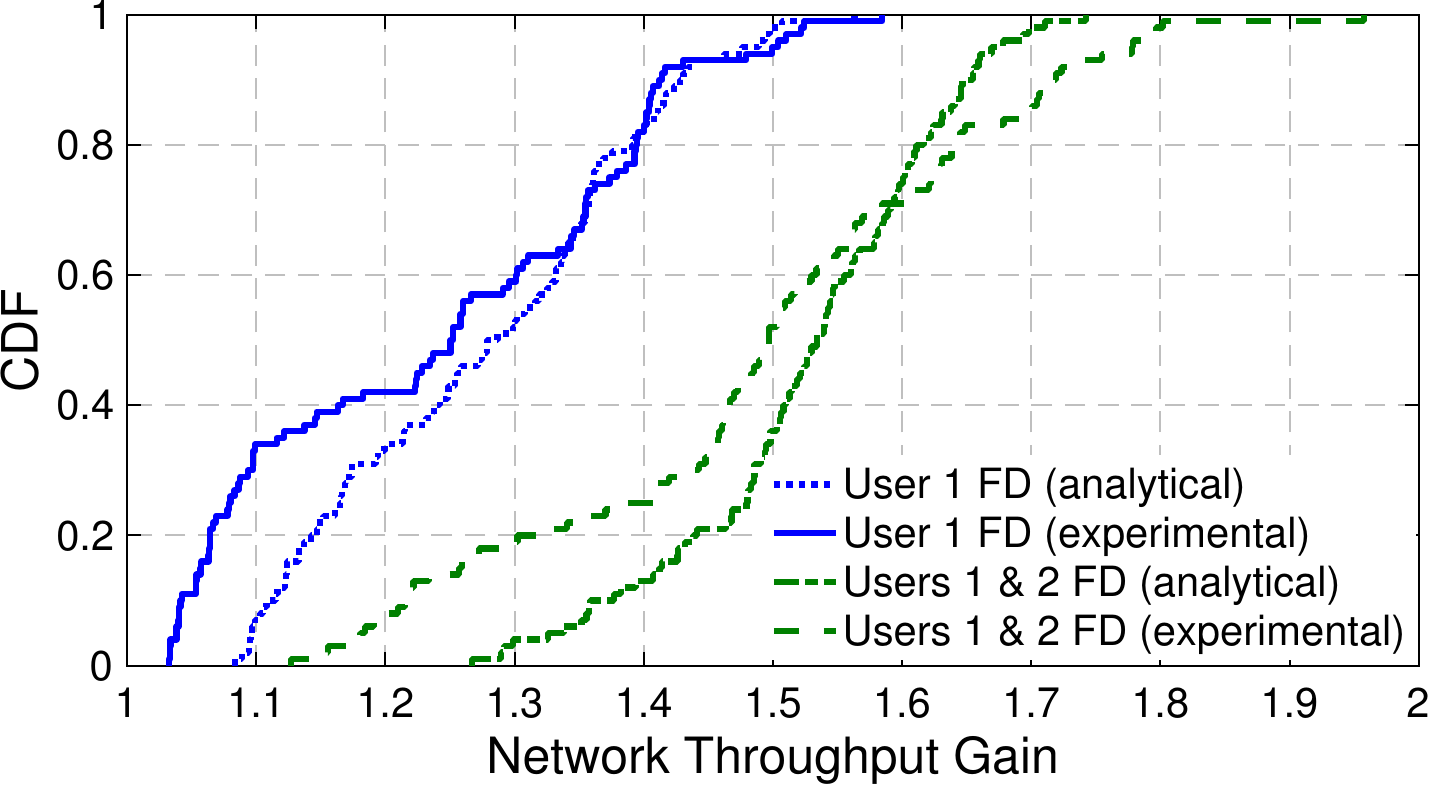}
\vspace{-0.5\baselineskip}
\caption{CDF of the analytical and experimental network throughput gains in 4-node networks when one or two users are FD-capable.}
\label{fig:exp-net-three-users-analysis-vs-exp}
\vspace{-\baselineskip}
\end{figure}

Fig.~\ref{fig:exp-net-three-users} shows the CDF of the network throughput of the three cases, where the measured link SNR varies between {5--45}\thinspace{dB}.Overall, the median network throughput is increased by 1.25/1.52$\times$ when one/two FD users become FD-capable.
Fig.~\ref{fig:exp-net-three-users-fairness} plots the CDF of the corresponding JFI, where although introducing FD-capable users results in lower values of the experimental JFI, the median JFI values degrade by only 0.06 and 0.10 with one and two FD users, respectively. Moreover, Fig.~\ref{fig:exp-net-three-users-analysis-vs-exp} shows the CDF of both the analytical and experimental network throughput gains in 4-node networks when one or two users are FD-capable. In particular, the median analytical and experimental network throughput gains have a difference of only 4\% and 3\% when one and two users are FD-capable.
These trends and results show that in a real-world environment, the total network throughput increases as more users become FD-capable, and the improvement is more significant with higher user SNR values. Note that we only apply a TDMA scheme and a more advanced MAC layer (e.g.,~\cite{kim2013janus, chen2018hybrid}) has the potential to improve the FD gain and fairness performance. \addedMK{As with the 3-node experiment, a similar version is run on the COSMOS testbed in Section~\ref{sssec:cosmos-exp-tdma-4node}.}

\section{\addedMK{The COSMOS FD Radios}}
\label{ssec:cosmos}
\addedMK{We also implemented four improved FDE canceller PCBs in the open-access COSMOS testbed~\cite{raychaudhuri2020challenge}, so that they may be remotely accessed by the research community. In this section, we describe the testbed, as well as present experimental results for the UL-DL and heterogeneous TDMA network experiments. A step-by-step procedure on how to access and use the testbed is provided via tutorial~\cite{fd_tutorial_wiki} which makes use of the custom COSMOS server image \texttt{flexicon-cosmos.ndz}. Lastly, we describe an available dataset with traces to support the development of digital SIC algorithms.}

\subsection{\addedMK{Integration of the FDE Canceller}}
\label{sssec:cosmos-integration}
\addedMK{An initial version of the testbed integration was presented in~\cite{kohli2021open}. The COSMOS FD radios consist of a ``canceller box" containing the improved FDE canceller PCB as shown in Fig.~\ref{fig:gen2-integration}\subref{fig:gen2-integration-box}, which also contains a circulator and antenna tuner to support the single antenna interface. Four such boxes are mounted on the corners of a 4.5$\times$3.5\thinspace{m} rectangle within a square laboratory room (the same room housing the FD BS in Figs.~\ref{fig:exp-map}\subref{fig:exp-map-nlos} and~\ref{fig:exp-net-setup}), where they serve as the frontends to USRP X310 SDRs with SBX-120 RF daughterboards.}

\begin{figure}[!t]
\centering
\includegraphics[width=0.98\columnwidth]{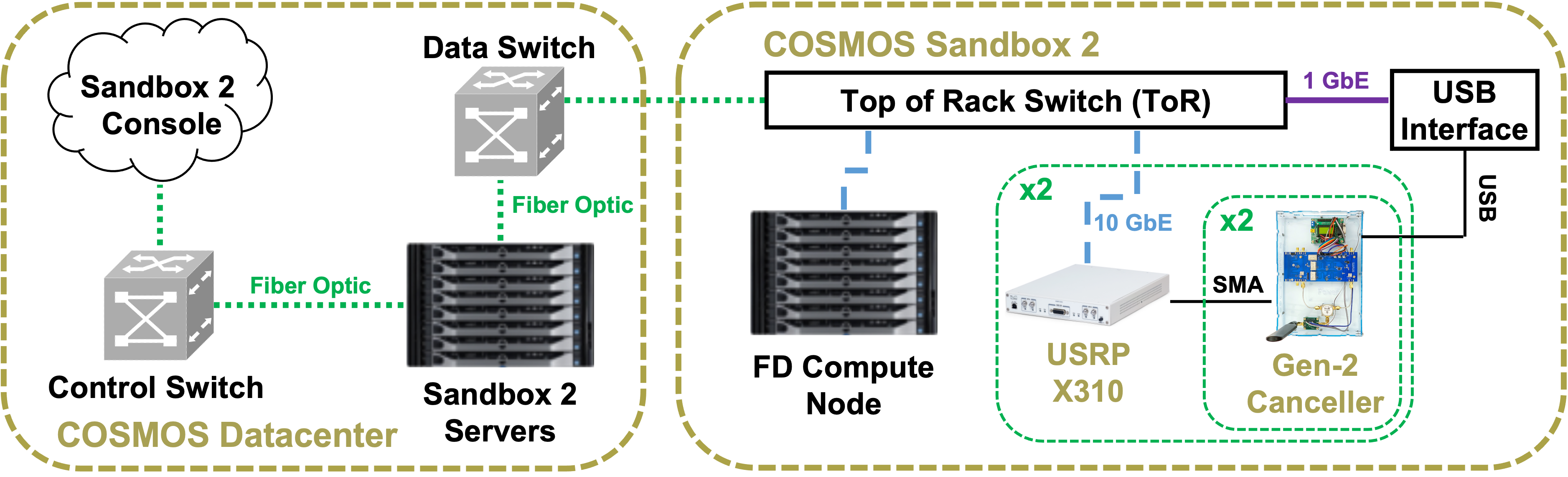}
\caption{\addedMK{Architecture of the FD testbed integrated in COSMOS Sandbox 2.}}
\label{fig:gen2-integration-arch}
\vspace{-\baselineskip}
\end{figure}

\addedMK{Two of the four radios are shown in Fig.~\ref{fig:gen2-integration}~\subref{fig:gen2-integration-rack}; the same setup with another two canceller boxes is duplicated on another ceiling rack directly opposite. Fig.~\ref{fig:gen2-integration-arch} shows an architecture diagram of the testbed. Individual parts of the testbed are detailed further below.}

\noindent\addedMK{\textbf{Improved PCB canceller.} The improved PCB's design follows the same outline provided in Section~\ref{ssec:impl-pcb}, but has been consolidated to include all components on a single PCB; the first revision requires a daughterboard to provide the analog voltage to control the FDE path phase shifters. The improved PCB canceller is configured manually via the experimentation software; the static laboratory environment housing the FD testbed leads to a time-invariant self-interference channel $\AntTF(f_k)$, allowing for stable canceller configuration. This also permits experimenters to generate various RF SIC profiles.}

\addedMK{Each of the four integrated improved FDE canceller PCBs can achieve over 50\thinspace{dB} of SIC across 20\thinspace{MHz} within the 900\thinspace{MHz} band~\cite{kohli2021open}. With digital SIC included, over 85\thinspace{dB} SIC may be achieved across 20\thinspace{MHz}. The canceller PCB design is publicly accessible, including the layout and bill of materials~\cite{flexicon_cosmos_github}.}

\noindent\addedMK{\textbf{USRP X310.} The prior iteration of the COSMOS FD testbed utilized a single USRP 2974 and two USRP N210s. The two types of SDR feature vastly different performance characteristics, predominantly caused by the different RF front ends. These radios were replaced with two USRP X310s with SBX-120 daughterboards, similar to the ones used in Section~\ref{ssec:exp-testbed}. The USRP X310s provide uniform performance and operation for all four radio nodes, and permit integration with the COSMOS servers via their 10\thinspace{Gb/s} interface.}

\noindent\addedMK{\textbf{Integration with COSMOS servers.} The largest improvement to the COSMOS FD testbed has been to connect the USRP X310 SDRs to two Dell R740 servers via 10\thinspace{Gb/s} links through a network switch. Each server is equipped with two Xeon 12-core CPUs and 192\thinspace{GB} of memory. An experimenter can log into and use these servers as the host PC to control the SDRs and canceller PCB for running experiments. A less powerful, dedicated-purpose FD Compute Node is also present within Sandbox 2 and may be used instead of the servers. We additionally utilize the servers or compute node to perform digital SIC for the experiments in Section~\ref{sssec:cosmos-exp}.}

\noindent\addedMK{\textbf{Experimentation software.} The lab FD testbed described in Section~\ref{ssec:exp-testbed} uses a LabVIEW-based software setup. For the COSMOS FD testbed, we migrated all experimental software to GNU Radio, facilitating the imaging process used for the COSMOS servers to save and reload experimental state. The experimental software is contained within a custom out-of-tree (OOT) module for GNU Radio~\cite{flexicon_cosmos_github}. This custom \texttt{C++} code supports real-time FD experimentation, including the digital SIC and MAC layer scheduling evaluated in this section.}

\subsection{Experimentation}
\label{sssec:cosmos-exp}
\addedMK{In this subsection, we describe the experiments conducted on the COSMOS FD radios.}

\subsubsection{UL-DL Networks with IUI}
\label{sssec:cosmos-exp-tdd}
\addedMK{We re-implemented the UL-DL network from Section~\ref{sssec:exp-net-ul-dl} to run on the COSMOS FD nodes. As these FD radios are fixed in place, the IUI from radio $A$ to another radio $B$ is a function of $A$'s transmit power only. This reduces the state space which can be explored in this version of the experiment compared to the version with freely-movable radios, where the IUI can be understood as a function of radio transmit power and position.}

\begin{figure}[!t]
\centering
\includegraphics[width=0.75\columnwidth]{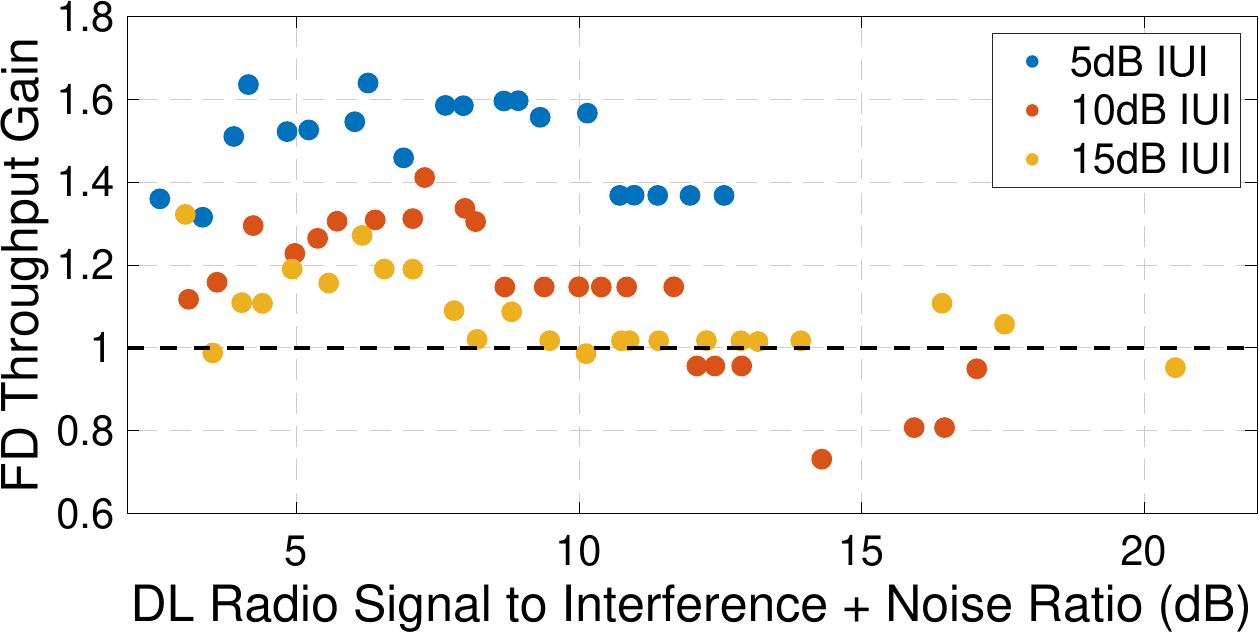}
\caption{\addedMK{FD throughput gains as a function of the DL radio SINR for IUI values of 5, 10, and 15\thinspace{dB}.}}
\label{fig:cosmos-exp}
\vspace{-\baselineskip}
\end{figure}

\addedMK{We present the experimental FD rate gains for the UL-DL network as achieved on the testbed integration in Fig.~\ref{fig:cosmos-exp}. We considered three IUI levels at the DL radio; 5, 10, and 15\thinspace{dB}. Using an 802.11-like PHY layer~\cite{bloessl2018performance}, we correspond these IUIs to specific MCS of QPSK 3/4, 16-QAM 1/2, and 16-QAM 3/4 respectively, which are used in the link between the UL and BS radios. We sweep the transmit power of the BS radio to generate SINR values at the DL radio sufficient for BPSK 1/2, BPSK 3/4, and QPSK 1/2 MCS. This testbed setup is representative of a densely populated, low power network prone to IUI. We used the dedicated FD compute node to run these experiments; they may be run in an identical manner using a Sandbox 2 server.}

\addedMK{Fig.~\ref{fig:cosmos-exp} shows that throughput gains relative to the ideal HD case are achievable for all considered IUI values. However, the gain drops significantly as the IUI increases; at 5\thinspace{dB} IUI a gain of up to 1.6$\times$ is achievable, whereas for 10\thinspace{dB} IUI the gain is reduced to a maximum of 1.3$\times$. This effect matches the analytical surfaces shown in Fig.~\ref{fig:exp-net-ul-dl-gain} which show a decrease in theoretical FD gain as a function of the IUI at the DL radio. We note that the specific FD rate gains shown in this experiment are comparable to those achieved in Section~\ref{sssec:exp-net-ul-dl}, demonstrating the FD testbed's ability to produce reliable and repeatable results.}

\begin{figure*}[!t]
\centering
\vspace{-\baselineskip}
\subfloat[]{
\label{fig:exp-net-two-users-cosmos-gain}
\includegraphics[width=0.42\linewidth]{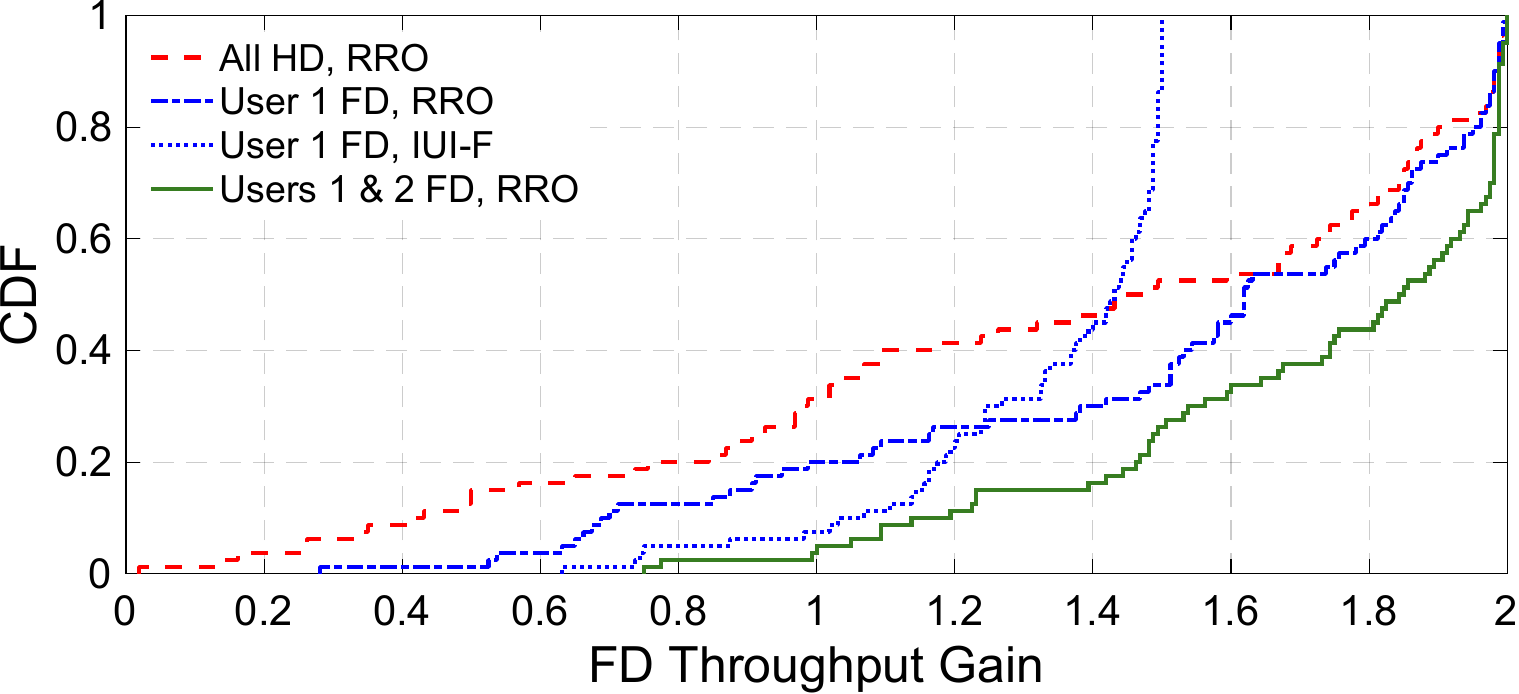}
}
\hspace{10pt}
\subfloat[]{
\label{fig:exp-net-two-users-cosmos-jain}
\includegraphics[width=0.42\linewidth]{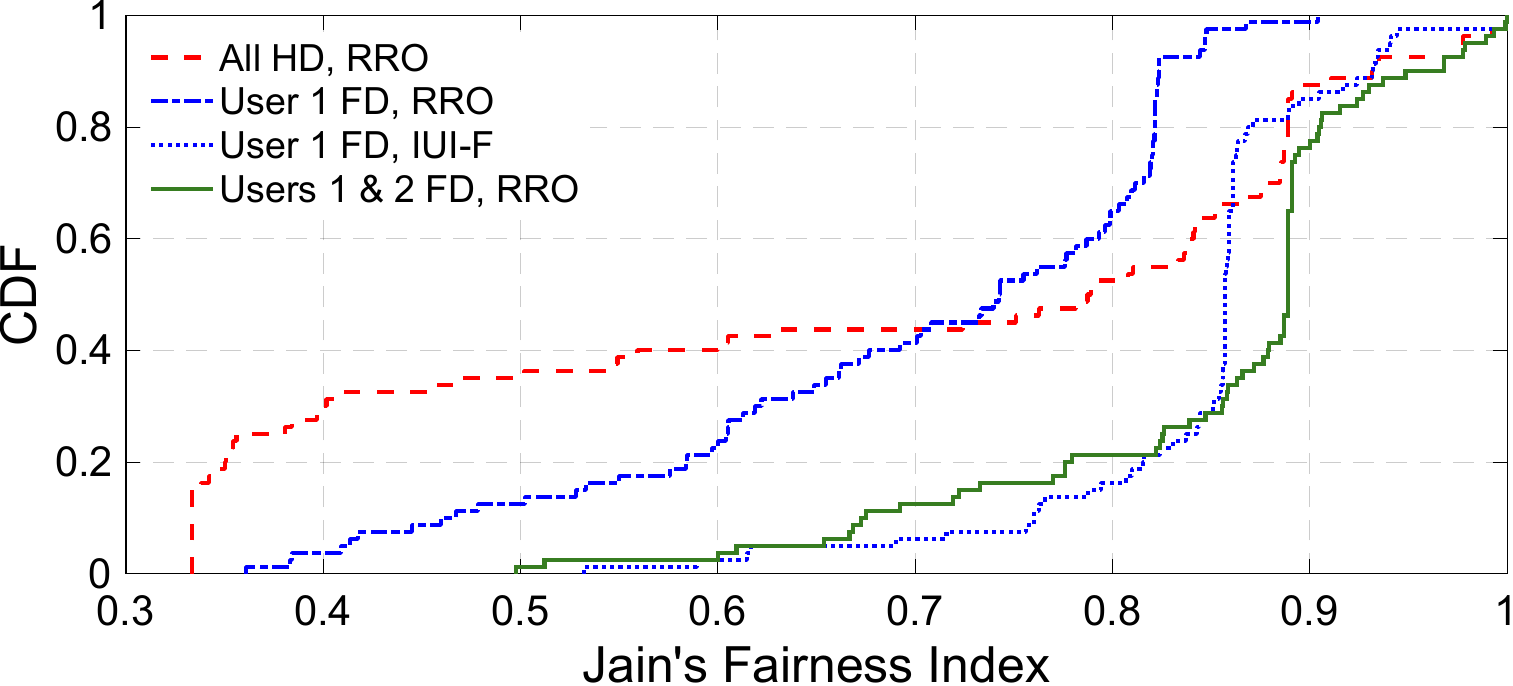}
}
\vspace{-0.5\baselineskip}
\caption{\addedMK{Experimental results for the heterogenous 3-node network. Results for (a) the FD throughput gain and (b) the Jain's fairness index are presented as CDFs. Zero, one, or two FD users (U) are considered, as well as the RRO and IUI-F schedules.}}
\label{fig:exp-net-two-users-cosmos}
\end{figure*}

\begin{figure*}[!t]
\centering
\subfloat[]{
\label{fig:exp-net-three-users-cosmos-gain}
\includegraphics[width=0.42\linewidth]{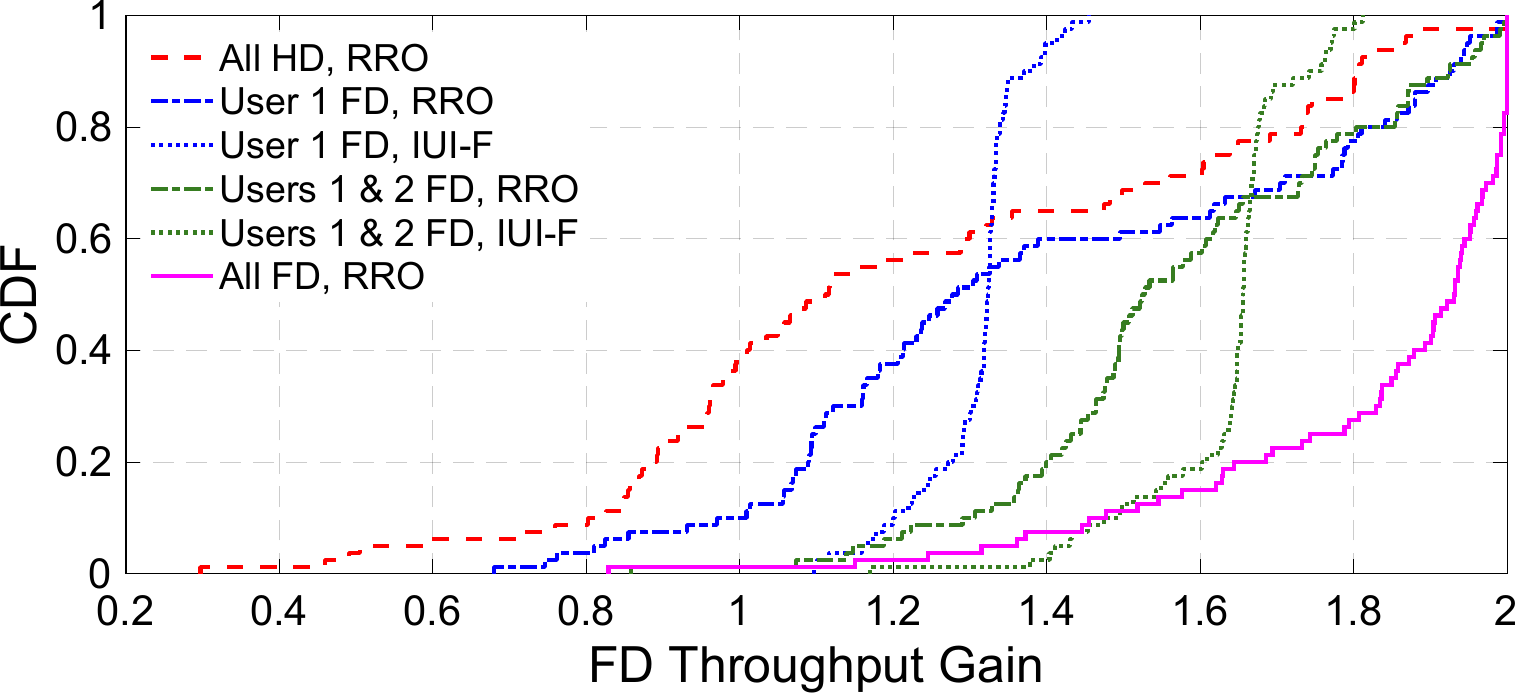}
}
\hspace{10pt}
\subfloat[]{
\label{fig:exp-net-three-users-cosmos-jain}
\includegraphics[width=0.42\linewidth]{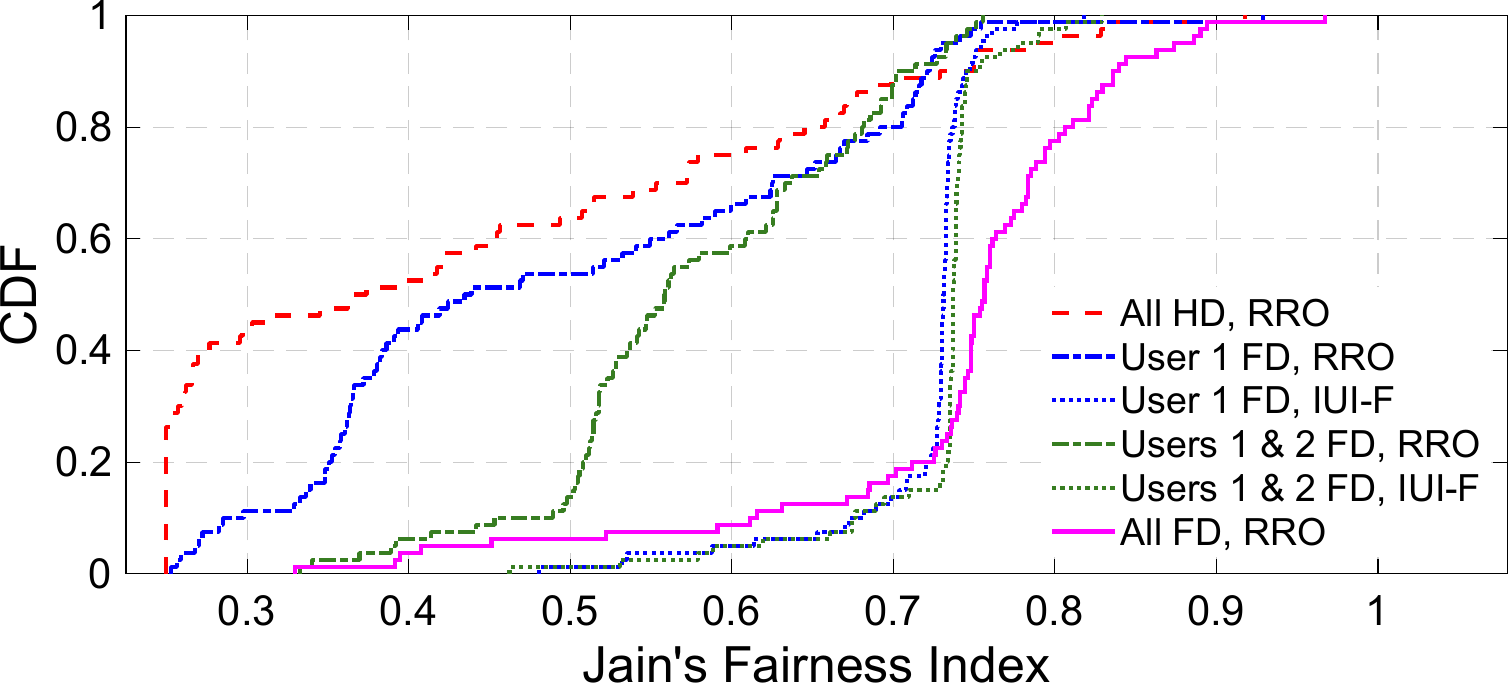}
}
\vspace{-0.5\baselineskip}
\caption{\addedMK{Experimental results for the heterogenous 4-node network. Results for (a) the FD throughput gain and (b) the Jain's fairness index are presented as CDFs. Zero to three FD users (U) are considered, along with the RRO and IUI-F schedules.}}
\vspace{-\baselineskip}
\label{fig:exp-net-three-users-cosmos}
\end{figure*}

\subsubsection{Heterogeneous 3-Node Networks}
\label{sssec:cosmos-exp-tdma-3node}

\addedMK{The experiment described in Section~\ref{sssec:exp-net-two-users} is also replicated on the COSMOS nodes. We consider two TDMA schedules; the first is identical to the one used in Section~\ref{sssec:exp-net-two-users}, henceforth referred to as the ``round-robin opportunistic" (RRO) schedule, and the second additional schedule is IUI-free (IUI-F). The RRO schedule works by opportunistically scheduling two links whenever possible in a fair, round-robin manner; if there are users operating in HD mode, there will necessarily be IUI as the FD BS will attempt to transmit to a different user. For example, IUI will occur in a TDMA slot where HD user 1 transmits to the FD BS, and the BS simultaneously transmits to user 2. In the IUI-free (IUI-F) schedule, this type of schedule is not permitted; if two radios are scheduled in the same time slot, they must both be in FD mode. Intuitively, the IUI-F schedule should maintain a higher throughput fairness between the users, at the expense of FD throughput gain.

In order to generate a range of SINRs at each radio, we sweep the transmit power of the BS across a 10\thinspace{dB} range, such that in HD mode, all packets may be received from the BS by the users. The transmit power of the users is fixed, and also chosen such that for any MCS, all packets may be received by the BS in HD mode. We consider a range of MCS between BPSK 1/2 and 64-QAM 3/4 available in the 802.11-like PHY layer~\cite{bloessl2018performance}; all radios use the same MCS.}

\addedMK{The results of this experiment are summarized in Fig.~\ref{fig:exp-net-two-users-cosmos}. Fig.~\ref{fig:exp-net-two-users-cosmos}\subref{fig:exp-net-two-users-cosmos-gain} shows the CDF of FD throughput gain for the different schedules, and Fig.~\ref{fig:exp-net-two-users-cosmos}\subref{fig:exp-net-two-users-cosmos-jain} shows the Jain's fairness index. We make two observations; the IUI-F schedule with zero FD users is equivalent to the HD TDMA schedule, and the IUI-F schedule with two FD users is equivalent to the RRO schedule. These two cases are therefore omitted from the figures. Fig.~\ref{fig:exp-net-two-users-cosmos}\subref{fig:exp-net-two-users-cosmos-gain} shows how having zero FD users can impair the RRO schedule significantly; this is likely due to the additional SNR requirement in FD mode~\cite{chen2019wideband, kohli2021open}. This effect is still observed when one or two users are in FD mode, but with reduced severity. Nevertheless, the three RRO schedules provide median throughput gains of 1.43$\times$, 1.6$\times$, and 1.85$\times$ when zero, one, or two users are in FD mode, respectively. By comparison, the IUI-F schedule for one FD user has a median throughput gain of 1.43$\times$.}

\addedMK{Fig.~\ref{fig:exp-net-two-users-cosmos}\subref{fig:exp-net-two-users-cosmos-jain} gives the median Jain's fairness index for the RRO schedule as 0.79, 0.74, and 0.89 for zero, one, or two FD users, respectively. The Jain's fairness index is lower with one FD user present on account of the ability to schedule the FD user whenever the HD user is scheduled, leading to the FD user receiving a higher throughput compared to the HD user. This is also the reason the one FD user experiment never reaches a fairness value of 1. The \mbox{IUI-F} schedule for one FD user has a median fairness of 0.86, showing how the fairness can be improved at the cost of FD throughput gain.}

\addedMK{The results from this experiment suggest than in a scenario where with high IUI between users, an RRO schedule can provide a 1.6$\times$ median FD throughput gain, but with the potential of low fairness between users. If an IUI-F schedule is used, fairness can be improved, though the median and maximum FD throughput gains will be lower. We do note that 20\% of experiments with one FD user on the RRO schedule experience a FD throughput \textit{loss}, compared to only 8\% of IUI-free schedule experiments. The IUI-free schedule therefore leads to a distribution of FD throughput gains with lower mean, but also lower variance.}

\subsubsection{Heterogeneous 4-Node Networks}
\label{sssec:cosmos-exp-tdma-4node}
\addedMK{A similar experiment, also considering the RRO and IUI-F schedules, was conducted with an additional FD-capable user. The corresponding results are presented in Fig.~\ref{fig:exp-net-three-users-cosmos}. In Fig.~\ref{fig:exp-net-three-users-cosmos}\subref{fig:exp-net-three-users-cosmos-gain}, the RRO schedule provides FD throughput gains of 1.11$\times$/1.28$\times$/1.52$\times$/1.93$\times$ when used with zero/one/two/three FD users. We again note that regardless of how many users are FD capable, the transmit power of each user is kept the same, and the transmit power of the BS is swept in the same way. Therefore, by simply running additional radios in FD mode, the median throughput gain increases. In other words, the median energy per bit remains the same, while the median throughput gain is improving.}

\addedMK{The IUI-F schedules provide an FD throughput gain of 1.32$\times$ and 1.66$\times$ when one or two users are in FD mode, respectively. Unlike the 3-node experiment in Section~\ref{sssec:cosmos-exp-tdma-3node}, the median FD throughput gain for the IUI-F schedule is actually higher than for the RRO schedule, although the maximum achievable FD throughput gain is naturally lower for the IUI-F schedule. The results show that when two users are FD capable, only 20\% of RRO experiments had a larger gain than the best IUI-F experiment.}

\addedMK{Fig.~\ref{fig:exp-net-three-users-cosmos}\subref{fig:exp-net-three-users-cosmos-jain} shows that the fairness value for the one and two FD user cases improves when the IUI-F schedule is used. For one FD user, the median fairness improves from 0.43 to 0.73, and for two FD users it improves from 0.56 to 0.74. Therefore, the 4-node experiment is in a situation where the IUI-F schedule provides an advantage for both median throughput and median fairness. This result suggests that when developing a MAC protocol for FD, the IUI can play a significant hand in the performance, and scheduling both HD and FD users in the same time slot of a TDMA system may lead to impaired performance should the IUI between the HD and FD users be high.}

\addedMK{The heterogeneous 4-node experiment is provided on the \texttt{flexicon-cosmos.ndz} COSMOS server image, and directions on how to run this experiment and reproduce the results in this section are provided in the testbed tutorial~\cite{fd_tutorial_wiki}.}

\subsection{Dataset}
\label{sssec:cosmos-exp-dataset}
\addedMK{We provide a dataset which consists of transmitted data packet waveforms and the corresponding received SI after RF SIC on the COSMOS dataset repository~\cite{cosmos_fd_dataset}. The data packets are QPSK 3/4-modulated 802.11a packets, similar to those used in the experiments presented in Sections~\ref{sec:exp} and \ref{ssec:cosmos}\footnote{Data for packets using other MCS may be collected from the COSMOS FD testbed}. The received power after RF SIC is between -40 to -50\thinspace{dBm} and the USRP noise floor is -80\thinspace{dBm} at 20\thinspace{MHz} bandwidth. 

The data files are in binary format representing the complex-valued digital baseband signals sent and received by the USRP SDR. In other words, this data is tapped after the ADC and before the DAC on the Rx and Tx paths within the NI USRP block in Figure~\ref{fig:diagram}. Every experimental run is represented by two data files, one each for the Tx and Rx baseband data, and every run contains at least 100 data packets. There are a total of twelve experimental runs across three bandwidths (5\thinspace{MHz}, 10\thinspace{MHz}, and 20\thinspace{MHz}) using each of the four radios as described in Section~\ref{sssec:cosmos-integration}. A document describing the experimental parameters in further detail is provided alongside the baseband data files.

The objective of this dataset is to provide ready-to-use baseband data traces to support the development of FD-related DSP algorithms, for example those for digital SIC. We provide an example MATLAB script that implements the same linear digital SIC algorithm as used in the testbed experiments in Section~\ref{sssec:cosmos-exp}. Furthermore, the GNU radio flowgraph used for data collection is provided on the \texttt{flexicon-cosmos.ndz} COSMOS server image. Further data files may be recorded from this experimental flowgraph with different parameters.}

%% file: tex/conclusion.tex
%


\addedMK{In this paper, we presented the system-level design, implementation, and evaluation of FDE-based FD radios based on~\cite{Zhou_WBSIC_JSSC15}. Experimentation conducted on two testbeds shows that that such radios, given an appropriate configuration optimization algorithm or a sufficiently stable physical environment, will achieve rate gains and overall performance that matches closely to analytical results. At the canceller design level, future research directions could include (i) better design and implementation of FDE-based canceller to support higher TX power handling and RF SIC bandwidth and (ii) extension of the FDE technique to multi-antenna systems.}

\addedMK{The COSMOS FD testbed is openly accessible to researchers, with a tutorial~\cite{fd_tutorial_wiki} describing its use. We anticipate that the COSMOS FD testbed will be used to explore several directions of future research, including the development and experimental evaluation of resource allocation and scheduling algorithms tailored for FDE-based FD radios. Additionally, the server-focused architecture of the COSMOS FD testbed effectively provides an edge compute environment, enabling future investigation into how FD will make use of this key component of next-generation cellular networks~\cite{smida2023fullduplex}. Lastly, we will integrate newer, wider-band time-domain equalization-based RFIC cancellers in the COSMOS FD testbed~\cite{levin2023experimentation,nagulu2021fullduplex}, as well as perform FD experimentation with the outdoors COSMOS testbed infrastructure~\cite{raychaudhuri2020challenge}.}